\documentclass[preprint, journal]{vgtc}              % review (journal style)
\onlineid{2057}

\preprinttext{%
To appear in IEEE Transactions on Visualization and Computer Graphics.}
\ieeedoi{xx.xxxx/TVCG.201x.xxxxxxx}

\vgtccategory{Theoretical \& Empirical}

\title{Visual Indicators to Increase the Detection of Linguistic Media Bias}
\renewcommand{\and}{\texorpdfstring{%
  \end{tabular}\hskip 1.5em  \begin{tabular}[t]{c}%
}{, }}

\author{%
  \authororcid{Smi Hinterreiter}{0000-0002-7029-2753}
  \and
  \authororcid{Anna Chelsea Bahß}{0009-0006-7003-1197}
  \and
  \authororcid{Ann-Christin Gah}{0000-0002-5634-625X}
  \and
  \authororcid{Timo Spinde}{0000-0003-3471-4127}
  \and
  \authororcid{Isao Echizen}{0000-0003-4908-1860}
  \and
  \authororcid{Marc Erich Latoschik}{0000-0002-9340-9600}
}

\authorfooter{
  \item
  	Smi Hinterreiter is with University of Würzburg.
  	E-mail: smi.hinterreiter@uni-wuerzburg.de.
  \item
  	Anna Chelsea Bahß is with University of Konstanz.
  	E-mail: a.bahss@online.de.
    \item
  	Ann-Christin Gah is with Technical University of Munich.
  	E-mail: ann-christin.gah@tum.de.
    \item
  	Timo Spinde is with  University of Göttingen and National Institute of Informatics Japan.
  	E-mail: t.spinde@media-bias-research.org.
    \item
  	Isao Echizen is with National Institute of Informatics Japan.
  	E-mail: iechizen@nii.ac.jp.
    \item
  	Marc Erich Latoschik is with University of Würzburg.
  	E-mail: marc.latoschik@uni-wuerzburg.de.
}

\abstract{%
The influence of linguistic bias in online news articles is a growing concern, particularly in the context of shaping public opinion and rising political polarization.
While there is a growing body of literature on indicators for misinformation, none have been sufficiently tested to counteract the influence of media bias.
Hence, we design six indicators (Bias Bar, Bias Gauge, Bias Highlights, Political Scale, Sentiment Scale, and Trust Score) and test their impact on linguistic bias detection and perception in a two-phased experiment (n = 214).
First, we expose participants to short, social-media-like statements along with one indicator and query bias perception.
Second, we evaluate bias detection by removing the indicator and asking participants to mark biased words.
In addition, we examine how trust, sharing discernment, and sentiment relate to bias perception and detection.
Our results show that highlighting biased phrases and showing total bias with contextual information in a gauge significantly improve bias detection skills.
However, the strongest predictor for reduced bias detection was political congruency between the statement and the participant.
We conclude with design recommendations for linguistic media bias indicators in online news environments.
}


\teaser{
  \centering
  \begin{subfigure}[b]{0.5\textwidth}
    \centering
    \includegraphics[width=\textwidth, alt={Example sentence with biased phrases highlighted in gray to illustrate text-based bias highlighting. Sentence says 'Both legal and illegal aliens will occupy 75 percent of new American jobs in as little as five years.' The phrases 'legal and illegal aliens' and 'as little as' are highlighted.}]{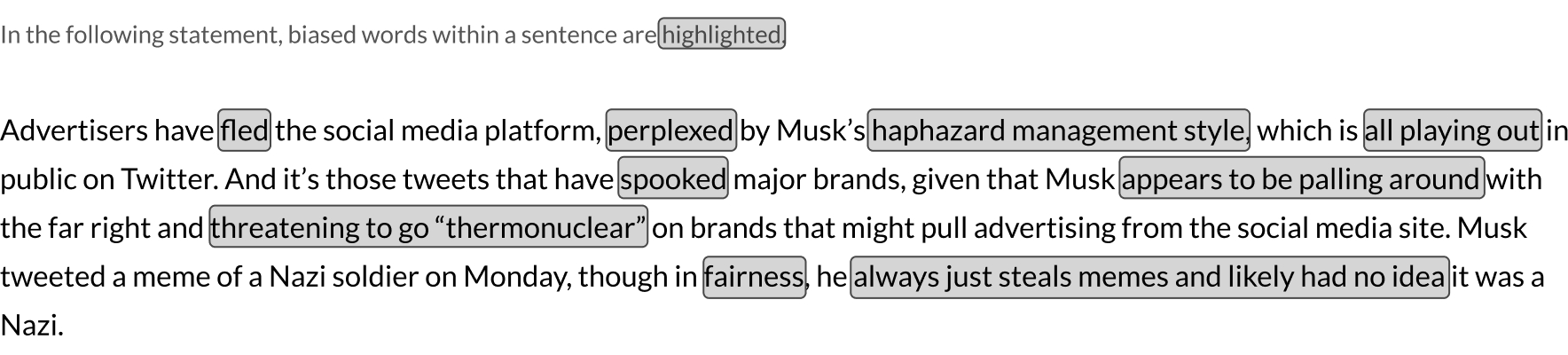}
    \caption{Bias Highlights. Positive Effects.}
    \label{fig:indicator_highlights}
  \end{subfigure}%
  \hspace{24pt}%
  \begin{subfigure}[b]{0.148\textwidth}
    \centering
    \includegraphics[width=\textwidth, alt={Semi-circular bias gauge with three sections (first 10\%, 10-30\%, and over 30\%) a needle pointing to the medium bias section and the label 'Medium Bias' below.}]{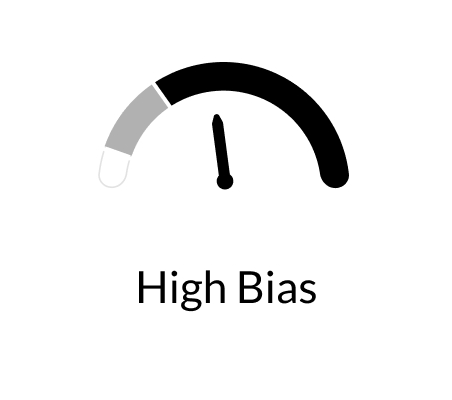}
    \caption{Bias Gauge. Positive Effects.}
    \label{fig:indicator_gauge}
  \end{subfigure}%
  \hspace{24pt}%
  \begin{subfigure}[b]{0.2\textwidth}
    \centering
    \includegraphics[width=\textwidth, alt={Horizontal bar labeled “Amount of Bias,” with the left portion filled around 25\% to indicate a moderate level of bias.}]{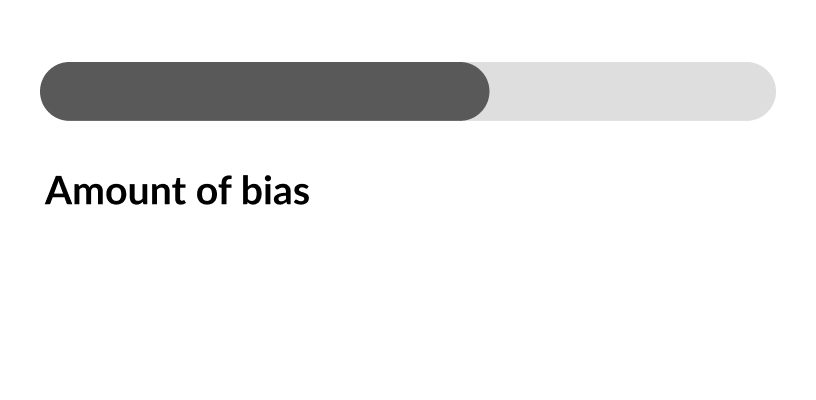}
    \caption{Bias Bar. Negative Effects.}
    \label{fig:indicator_bar}
  \end{subfigure}%
  \\[3pt]%
  \begin{subfigure}[b]{0.27\textwidth}
    \centering
    \includegraphics[width=\textwidth, alt={5-Point political leaning scale from Left to Right with according political labels and a slider positioned at the far-right end.}]{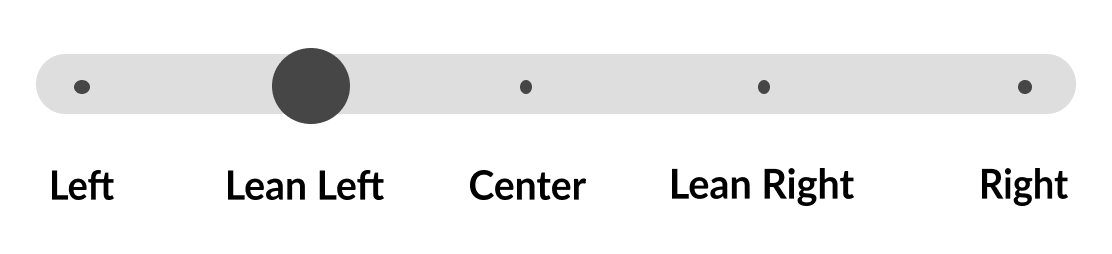}
    \caption{Political. No Effect.}
    \label{fig:indicator_pol}
  \end{subfigure}%
  \hspace{24pt}%
  \begin{subfigure}[b]{0.25\textwidth}
    \centering
    \includegraphics[width=\textwidth, alt={Horizontal sentiment indicator named Emotionality with a partially filled bar and labels for negative, positive, and neutral sentiment categories.}]{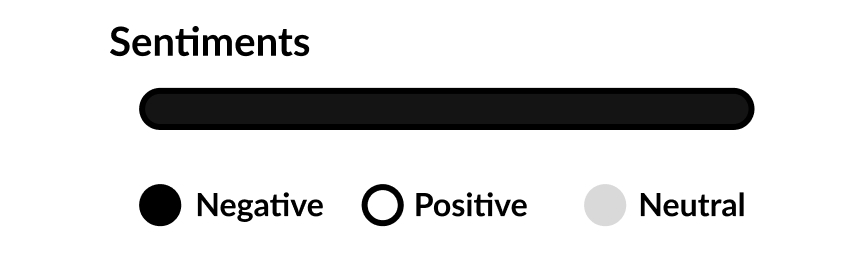}
    \caption{Emotionality. No Effect.}
    \label{fig:indicator_emo}
  \end{subfigure}%
  \hspace{24pt}%
  \begin{subfigure}[b]{0.32\textwidth}
    \centering
    \includegraphics[width=\textwidth, alt={Trust indicator showing two of six shield icons filled, labeled '2 of 6 trust points.'}]{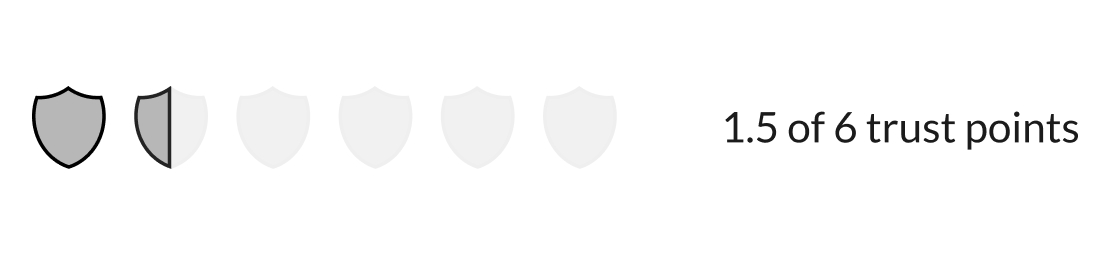}
    \caption{Trust. Positive Tendency.}
    \label{fig:indicator_trust}
  \end{subfigure}%
  \caption{The six tested indicators on one example statement from the study. Top row from left to right: \textit{Bias Highlights}, \textit{Bias Gauge}, and \textit{Bias Bar}. Bottom row from left to right: \textit{Political}, \textit{Emotionality}, and \textit{Trust}.}
  \label{fig:indicators}
}

\graphicspath{{figs/}{figures/}{pictures/}{images/}{./}} % where to search for the images

\usepackage{tabu}                      % only used for the table example
\usepackage{booktabs}                  % only used for the table example
\usepackage{lipsum}                    % used to generate placeholder text
\usepackage{mwe}                       % used to generate placeholder figures
\usepackage{ccicons}                   % package to be able to use icons from

\usepackage{mathptmx}                  % use matching math font
\begin{document}

%%%%%%%%%%%%%%%%%%%%%%%%%%%%%%%%%%%%%%%%%%%%%%%%%%%%%%%%%%%%%%%%
%%%%%%%%%%%%%%%%%%%%%% START OF THE PAPER %%%%%%%%%%%%%%%%%%%%%%
%%%%%%%%%%%%%%%%%%%%%%%%%%%%%%%%%%%%%%%%%%%%%%%%%%%%%%%%%%%%%%%%

\hypersetup{
  pdftitle={Visual Indicators to Increase the Detection of Linguistic Media Bias},
  pdfauthor={Smi Hinterreiter, Anna Chelsea Bahss, Ann-Christin Gah, Timo Spinde, Isao Echizen, Marc Erich Latoschik}
}

\pdfstringdefDisableCommands{%
  \def\authororcid#1#2{#1}%   % for PDF-string purposes, just use the name, drop the ORCID icon/link
  \def\and{, }%               % treat \and as a comma when flattening to plain text
}
\firstsection{Introduction}

\maketitle
%% \section{Introduction} %for journal use above \firstsection{..} instead

News articles often contain bias toward political ideologies, groups, or narratives \cite{leeNeuSNeutralMultiNews2022}.
This media bias can influence public perception and voting behavior \cite{effectsArd2017, leeNeuSNeutralMultiNews2022}.
Perceived bias depends on readers’ backgrounds \cite{effectsArd2017} and can limit exposure to diverse viewpoints \cite{FramingClimateUncertaintyGaissmaier2019}.
Media bias has multiple dimensions: selection bias \cite{Alonso_2017, fairBudak206} determines coverage, framing \cite{Recasens_2013} or spin bias \cite{Mullainathan_2002} shapes presentation, and linguistic bias \cite{Hube_2018b, spindeNeuralMediaBias2021} signals viewpoints and evokes emotions through word choice \cite{spinde2023media}.
We focus on linguistic bias, which often goes undetected by readers \cite{spindeEnablingNewsConsumers2020} but can be identified by automated methods \cite{Hube_2018b, Wessel2023}.

Despite its prevalence, research on media bias mitigation remains sparse \cite{spinde2023media}.
First studies on visual indicators, such as graphical cues that communicate bias or credibility at a glance \cite{Ware2004InformationVisualizationPerception}, showed promising results by making linguistic bias visible \cite{spindeHowWeRaise2022, simpleFramingInterventionBaumer2015}.
Yet only three studies specifically tested visual indicators in the context of media bias.
Among tested approaches such as warning messages and political classifications, highlighting biased phrases has shown the most consistent results \cite{spindeHowWeRaise2022, spindeEnablingNewsConsumers2020, simpleFramingInterventionBaumer2015}.
Broader work on misinformation and news credibility indicators offers more evidence on which design strategies for visual indicators might help critical evaluation of information \cite{Bodaghi2024}.
However, prior work is difficult to compare: studies diverge in their metrics and content, and most test only one indicator at a time \cite{simpleFramingInterventionBaumer2015}. Critically, most studies rely solely on self-reported perception, without active detection tasks that could reveal whether indicators genuinely improve readers' ability to identify bias \cite{spindeHowWeRaise2022}. Further, not all study designs account for political leaning of participants and congruency effects \cite{sultanSusceptibility2024}.

This study addresses these gaps by identifying visual indicators for linguistic bias, then designing and comparing six indicator types using standardized measures and a detection task.
Indicator type serves as the treatment variable, with a controlled first-exposure design testing transfer effects for linguistic bias and related concepts after exposure.
We follow these research questions:

\begin{itemize}
    \item \textbf{RQ1:} Which visual indicators increase linguistic bias detection?
    \item \textbf{RQ2:} Which visual indicators increase self-reported bias perception?
    \item \textbf{RQ3:} How do sharing intention, perceived emotionality, and trust relate to perceived bias?
    \item \textbf{RQ4:} How does political congruency moderate bias perception and bias detection?
\end{itemize}

\begin{figure*}[!ht]
  \centering 
  \includegraphics[width=1.5\columnwidth, alt={The figure is a left-to-right flowchart with three tall rounded rectangles connected by arrows across the top. Each panel has a black header with white text: “(1) Intro,” “(2) Indication Phase,” and “(3) Testing Phase.” In the first panel, the study begins with “Consent to Data Processing (n = 241),” followed by downward arrows to “Demographics Questionnaire,” then “Explanation of Media Bias + Examples,” then “Attention Check No. 1 (n = 233).” Below that, participants are “Randomly Assigned to one of the 7 Groups,” listed as Bias Bar, Bias Highlights, Bias Gauge, Political, Emotional, Trust, and Control. The second panel begins with “Indicator Condition Introduction.” A downward arrow leads to a repeated task block labeled “Statement + Indicator Condition + Question Items,” accompanied by curved arrows on both sides to show repetition and the note “Repeated for 9 Statements.” At the bottom of this panel is “Attention Check No. 2 (n = 233).” The third panel begins with “Bias Annotation Task Introduction.” A downward arrow leads to another repeated task block labeled “Statement + No Indicator + Question Items + Bias Annotation,” again marked with curved arrows and the note “Repeated for 6 Statements.” Below that, the figure states: “Check if min. one word was marked biased if participant perceived bias (n = 221),” then “Data Usable for Research? (n = 214),” and finally “Thank You + Debrief.”}]{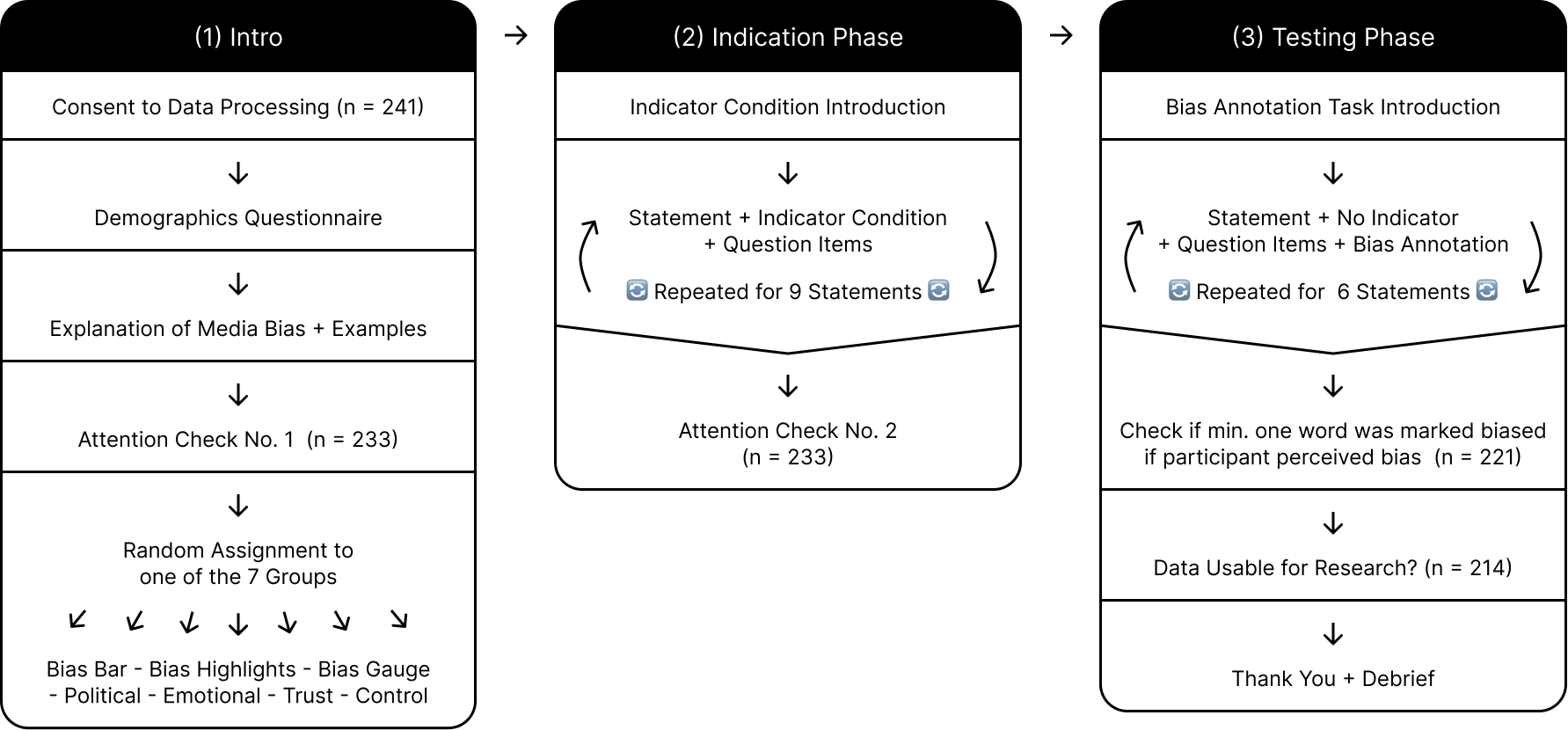}
  \caption{%
  	Study procedure with the three phases (1) Intro, (2) Indication Phase, and (3) Testing Phase, each with measures, steps, and included participants.
  }
 \label{fig:process}
\end{figure*}

To create the indicators, we first review related work from media bias, misinformation, and information credibility research (\Cref{sec:rw:design}) to create a design framework (\Cref{sec:meth:indicators}).
We then design six indicators, shown in \Cref{fig:indicators}, covering bias amount (\textit{Bias Bar}, \textit{Bias Gauge}, \textit{Bias Highlights}), political leaning (\textit{Political}), sentiment (\textit{Emotionality}), and trustworthiness (\textit{Trust}), and compare them against a control group in an online study (n = 214). The study focuses on short, social-media-like statements and includes both a bias perception measure and a word-level annotation test for evaluation beyond self-report (\Cref{sec:meth:measures}). In two phases, participants first read statements with their group's indicator before moving to a testing phase where they complete a bias detection test without their group's indicator (\Cref{fig:process}).

Our results show that \textit{Bias Highlights} and, partly, \textit{Bias Gauge} significantly improve detection accuracy (\Cref{sec:results}).
However, conservative or liberal political leaning and incongruence with a statement's slant are strong predictors of both detection and perception across all groups.
We also find a strong correlation between perceived emotionality and perceived bias.
We contribute an empirical comparison of visual representation strategies for linguistic bias indicators, evidence that phrase-level and contextualized aggregate encodings outperform abstract summary cues for transfer-based bias detection, and design implications about interpretability, reference frames, and calibration for reflective bias judgment tools under political congruence (\Cref{sec:dis}).\footnote{Code and data available at \url{doi.org/10.5281/zenodo.19347505}}

\section{Related Work}
Prior work suggests that making bias visible can raise awareness and mitigate negative effects \cite{simpleFramingInterventionBaumer2015, An_Cha_Gummadi_Crowcroft_Quercia_2021}.
Among tested approaches for media bias are highlighting facts in the text, highlighting instances of framing, highlighting biased language \cite{spindeEnablingNewsConsumers2020, simpleFramingInterventionBaumer2015}, forewarning messages, and political maps \cite{An_Cha_Gummadi_Crowcroft_Quercia_2021} or scales \cite{spindeHowWeRaise2022}.
So far, the highlighting of biased phrases (see \Cref{fig:examplesinterventions}) emerged as the most promising approach and has been implemented in online applications \cite{Hinterreiter2025NewsUnfold}.
However, studies on bias mitigation remain limited, and only a small number directly test visual indicators for media bias perception \cite{spindeHowWeRaise2022, spindeEnablingNewsConsumers2020, simpleFramingInterventionBaumer2015}, with more evidence coming from related work on news indicators for misinformation mitigation and credibility assessment.

\subsection{Related Indicator Designs} \label{sec:rw:design}
To identify indicator designs for making linguistic bias visible, we reviewed 33 publications on news-related indicators, also including commonly used commercial systems such as NewsGuard \cite{NewsGuard2022}, The Factual, and AllSides.\footnote{\url{web.archive.org/web/20240813080330/https://www.thefactual.com/how-it-works/}, \url{www.allsides.com/}}.
The full table of the literature review is available in the supplementary material (\Cref{sec:supplemental_materials}).
Across included work, indicators differ in the strategy they use to indicate credibility or bias, timing of the indicator, and the data type displayed.
Regarding timing, some intervene before reading (e.g., forewarning messages \cite{thornhillDigitalNudgeCounter2019}), while others provide information during reading through labels, highlights, scales, or ratings (e.g., bias highlights \cite{spindeHowWeRaise2022}). Other interventions use debunking strategies \cite{Chan2017debunking} after reading or provide alternative sources \cite{spindeEnablingNewsConsumers2020}.
We further classify interventions into six categories based on the strategy of the intervention:
Manipulation of news presentation \cite{KirchnerCountering2020, spindeEnablingNewsConsumers2020};
Warning messages \cite{sys1sys2InterventionsMoravec2020, thornhillDigitalNudgeCounter2019};
Educational prompts or messages \cite{guessDigitalMediaLiteracy2020, sys1sys2InterventionsMoravec2020};
Social cues such as likes or endorsements \cite{luoCredibilityPerceptionsDetection2022}, \Cref{fig:examplesotherindicators};
Recommendations \cite{NewsGuard2022, kimCombatingFakeNews2019a, kimSaysWhoEffects2019, bhuiyanNudgeCredSupportingNews2021};
Information about article or source \cite{spindeHowWeRaise2022, spindeEnablingNewsConsumers2020, An_Cha_Gummadi_Crowcroft_Quercia_2021}, \Cref{fig:examplesinterventions}.
%Among tested bias-specific approaches, highlighting biased phrases has shown the most consistent effects on bias awareness \cite{simpleFramingInterventionBaumer2015, spindeEnablingNewsConsumers2020, spindeHowWeRaise2022} and has been implemented in pilot projects \cite{Hinterreiter2025NewsUnfold}.
Of the six strategies, information and recommendation indicators are used most frequently, especially by commercial applications.

\textbf{Recommendation Indicators} designed to assist users to evaluate the credibility of news can potentially be more effective in enhancing perceived credibility than fact-checking indicators \cite{bhuiyanNudgeCredSupportingNews2021}.
Examples include The Factual’s credibility score and NewsGuard’s red or green shield icons, although the latter did not demonstrate significant improvements in news consumption \cite{NewsGuard2022}.
Other work \cite{kimCombatingFakeNews2019a, kimSaysWhoEffects2019} indicates that source ratings can impact the perceived credibility of an article, with article ratings exerting a stronger influence on believability than source ratings.
Thus, recommendations present a conflicting research landscape, highlighting a need for further investigation.

\textbf{Information Indicators} present content-related cues outside the content itself to help readers assess its credibility and bias.
For example, the highlighting of instances of framing bias \cite{simpleFramingInterventionBaumer2015} and biased words and phrases \cite{spindeHowWeRaise2022} increases media bias awareness and reduces ideological polarization.
Platforms like AllSides often use political classifications to encourage readers to consider diverse perspectives.
However, ideological left-right maps did not significantly affect bias perception in the study of Spinde et al. \cite{spindeHowWeRaise2022} while other studies describe positive effects \cite{parkNewsCubeDeliveringMultiple2009}.

\subsection{Visual In-Text and Summary Encodings for Media Bias}
Highlighting biased phrases shows the most consistent effects on bias awareness \cite{simpleFramingInterventionBaumer2015, spindeEnablingNewsConsumers2020, spindeHowWeRaise2022}.
Text highlights act as in-text encodings that increase perceptual salience rather than summarizing content.
They function as pre-attentive popout cues whose effectiveness depends on density, contrast, and whether they encode categorical or continuous distinctions \cite{strobeltGuidelinesEffectiveUsage2016a, Ware2004InformationVisualizationPerception}.
Like overlays, highlights guide attention within text, unlike summary indicators such as bars, scales, or icons that present aggregated judgments.
In media bias contexts, highlights support cue localization and learning-by-example \cite{healeyAttentionVisualMemory2012}, which may enable transfer, while summary indicators support holistic assessment via heuristics \cite{franconeriScienceVisualData2021, sundarMAINModelHeuristic}.
Their effects on linguistic bias perception and detection have not been systematically compared.

Summary indicators aggregate bias signals into a single display outside the text.
Continuous bars provide accurate magnitude estimates \cite{Cleveland1984, Heer2010VisuDesign}, but without reference points, values are hard to interpret \cite{shahReviewGraphComprehension2002}.
Categorical segmentation adds labeled thresholds that reduce cognitive load in mapping values to judgments \cite{swellerCognitiveLoadDuring1988, franconeriScienceVisualData2021}.
Ordinal labels encode rank but not distance \cite{Mackinlay1986classic, munznerVisualizationAnalysisDesign2014}, so category steps lack quantitative meaning.
Composition displays (e.g., sentiment bars) increase complexity; accuracy declines with more components, and readers anchor on salient categories \cite{simkinInformationProcessingAnalysisGraph1987}.
Icon-based displays (e.g., stars, shields) act as heuristic cues that shape credibility judgments, often independent of the encoded score \cite{sundarMAINModelHeuristic, kimSaysWhoEffects2019, LixunTrustworthy2019}.

\subsection{Gaps in Indicator Studies}
%Existing indicator studies show several limitations. First, many compare a single indicator against a control group, which makes results hard to compare across studies \cite{sultanSusceptibility2024}. Second, media bias studies often rely only on self-reported bias perception, even though perception does not necessarily reflect actual detection ability \cite{spindeYouThinkIt2021}. Third, personal factors such as political leaning, age, and education are often not modeled, despite their known influence on news judgments \cite{sultanSusceptibility2024, spindeHowWeRaise2022}.
%Fourth, methodological differences in materials and measures limit comparability \cite{sultanSusceptibility2024}, such as different content formats, question items, or scales.
%For example, misinformation studies typically use news headlines and either ask for binary veracity of the presented information (true/false) or sharing discernment \cite{Bodaghi2024}.
%Conversely, media bias studies use articles or longer statements and ask for bias perception or topic opinions using scales \cite{simpleFramingInterventionBaumer2015, spindeHowWeRaise2022}.
Existing indicator studies have four main limitations: (1) most compare a single indicator to a control group, which makes results hard to compare across studies \cite{sultanSusceptibility2024}, (2) they often rely only on self-reported bias perception, which does not necessarily reflect actual detection ability \cite{spindeYouThinkIt2021}, (3) they frequently omit personal factors such as political leaning, age, and education despite their known influence on news judgments \cite{sultanSusceptibility2024, spindeHowWeRaise2022}, and (4) differences in materials and measures, such as content formats, items, and scales, limit comparability across misinformation and media bias research \cite{sultanSusceptibility2024}. Misinformation studies typically use headlines and ask for binary veracity judgments or sharing discernment \cite{Bodaghi2024}, whereas media bias studies use longer texts and measure bias perception or topic opinions on rating scales \cite{simpleFramingInterventionBaumer2015, spindeHowWeRaise2022}.

%We address these gaps with three contributions: (1) the first empirical comparison of six indicator types for linguistic bias covering commonly deployed strategies, including two indicators previously tested for media bias (\textit{Bias Highlights}, \textit{Political}), two novel summary designs (\textit{Bias Bar}, \textit{Bias Gauge}), and two adapted from credibility tools (\textit{Emotionality}, \textit{Trust}), (2) a transfer-based bias detection task as a new behavioral measure beyond self-report additional to bias perception, and (3) modeling demographic and political factors.
We address the gaps with three contributions: (1) the first empirical comparison of six linguistic-bias indicator types that reflect commonly deployed strategies, including two indicators previously tested for media bias (\textit{Bias Highlights}, \textit{Political}), two novel summary designs (\textit{Bias Bar}, \textit{Bias Gauge}), and two adapted from credibility tools (\textit{Emotionality}, \textit{Trust}), (2) a transfer-based bias detection task as a behavioral measure beyond self-reported bias perception, and (3) models that include demographic and political factors. 
%The study uses a transfer design, first exposing participants to an indicator to then remove it to estimate whether indicator exposure produces effects on unaided bias detection. This measures learning-like transfer rather than in-context assistance.
%Given the growing importance of short-form news consumption on social media \cite{liedkeAdults30Now}, we focus on short statements rather than full articles or only headlines.
Using a transfer design, we first expose participants to an indicator and then remove it to test unaided bias detection on new short, social-media-like statements, capturing learning-like transfer rather than in-context assistance \cite{liedkeAdults30Now}.
%This study focuses on indicator type as the treatment variable, not individual encoding parameters.
The treatment variable is indicator type rather than encoding parameters.
As exploring all possible design combinations is infeasible, we adopt a top-down approach, selecting indicator types that cover commonly deployed indicator strategies displayed during reading which aim to inform or recommend, rather than exploring individual design details or exhausting the full design space.
Findings should be interpreted as evidence about broader indicator strategies under first-exposure short-form conditions rather than to specific implementation choices.

\section{Method and Materials}
\label{sec:method}
\subsection{Study Overview}
We conducted a 7x3 mixed-design study with repeated measures, comparing six visual indicators with a control condition. The study had one between-subjects factor \textit{Indicator Group} (\textit{Bias Bar}, \textit{Bias Gauge}, \textit{Bias Highlights}, \textit{Political}, \textit{Emotionality}, \textit{Trust}, \textit{Control}) and the within-subject factor \textit{Political Slant} (left, center, right).
%s in the indication phase, \textit{Bias Level} (low, medium, high) and
The procedure consisted of three phases displayed in \Cref{fig:process}: (1) introduction to linguistic bias and attention check, the (2) indication phase, in which participants read statements with one assigned indicator, and the (3) testing phase, in which participants identified biased words without indicator support (\Cref{sec:meth:procedure}).

The study combined subjective perception measures and a more objective test (\Cref{sec:meth:measures}). We measured perceived bias, perceived emotionality, trust, and sharing intention after each statement, and we measured bias detection through a word-level annotation task in the testing phase. We designed the testing phase as a transfer test: after exposure to one representation type during reading, participants completed bias detection on new statements without visual support. The study and models also included political leaning, political congruence, age, gender, education, topic opinion, and topic relevance in the models to examine how personal background shaped indicator effects.

\subsection{Indicator Design}
\label{sec:meth:indicators}
Following visual encoding theory \cite{Heer2010VisuDesign, Mackinlay1986classic, Cleveland1984}, we used Munzner's nested model \cite{munznerNestedModelVisualization2009} as a design framework, extending it by our indicator review. We defined five design dimensions: (1) indicator strategy and goal (increasing bias detection and perception by showing information or giving recommendations), (2) indicator context, (3) data type abstraction (e.g., bias amount, biased words, sentiments), (4) visual concept and mental model fitting the data abstraction, and (5) visual design details and platform implementation.
Our study holds three dimensions constant: (1) indicator strategy and goal are use information indicators and a single indicator that extends on information by adding an element of recommendation, (2) indicator context is set do during reading, and (5) visual design details like size, placement, and grayscale are standardized.
The varying factors are (3) data type abstraction and (4) visual concept and mental models. 

Using this framework, we selected six indicators (\Cref{tab:indicators}, \Cref{fig:indicator_bar}–\Cref{fig:indicator_trust}), four representing common designs in prior work (\textit{Bias Highlights}, \textit{Political}) or deployed systems (\textit{Trust}, \textit{Emotionality}) and two new summary indicator designs for bias (\textit{Bias Bar}, \textit{Bias Gauge}).
Both bias summary indicators use bias amount as a continuous quantitative value that maps to filled length along a common scale, one of the most accurately perceived visual encodings \cite{Cleveland1984, Heer2010VisuDesign}.
\textit{Bias Gauge} adds categorical segmentation, providing a reference frame that reduces the cognitive demand of interpreting raw magnitudes \cite{swellerCognitiveLoadDuring1988, franconeriScienceVisualData2021, Ware2004InformationVisualizationPerception}.
The six indicators differ along data type (bias amount, political slant, sentiment, credibility) which constrains feasible visual forms and encoding types (in-text localized highlights vs. summary aggregate as ratios above the text as scales or gauges or credibility values encoded in icons like shields or stars), and cognitive role (example-based learning vs. heuristic cue vs. reference frames), making the comparison more complex than a pure encoding study and grounded in real-world deployment practices.
Visual design details and implementation such as size, placement, and color were standardized across indicators to ensure comparability, while acknowledging that some characteristics are inherently tied to the indicator concept.

\begin{table}[!ht]
\footnotesize
\caption{Summary of the six tested indicators.}
\label{tab:indicators}
\centering
\begin{tabular}{p{0.95\columnwidth}}
\toprule
\textbf{Indicator, Data Type, Encoding, Mechanism \& Expected Cognitive Role} \\
\midrule
\parbox[t]{\linewidth}{\textbf{Bias Bar:} Continuously displays amount of biased words. (\Cref{fig:indicator_bar})\\  \textbf{Data Type:} Bias Amount (Percentage of biased words; text-level). \\   \textbf{Encoding:} Following Harris et al. \cite{harrisSearchingDiversePerspectives} and Gao et al. \cite{GaoIntelligentInterface2017}, the \textit{Bias Bar} shows the percentage of biased words in the statement (s. \Cref{sec:method:material}) through a dark fill, excluding stop words. \textbf{Mechanism:} Raw quantitative summary. \textbf{Cog Role:} Compact magnitude cue.} \\\hline

 \parbox[t]{\linewidth}{ \textbf{Bias Gauge:} Continuously displays percentage of biased words and categorizes the level of bias in low/medium/high. (\Cref{fig:indicator_gauge})\\  \textbf{Data Type:} Bias Amount + Frame (Percentage of biased words; text-level). \\  \textbf{Encoding:} Drawing inspiration from Kherwa et al. \cite{kherwaApproachComprehensiveSentimental2014}, the \textit{Bias Gauge}, similar to the \textit{Bias Bar}, fills up continuously to represent bias amount. It is divided into three sections ('low bias' for bias percentages below 10\%, 'medium bias' between 10\% and 30\%, and 'high bias' for over 30\%) with corresponding text below. The thresholds were established after analyzing the percentage values in statements categorized as low, medium, and high bias during material collection (\Cref{sec:method:material}). The \textit{Bias Gauge} aims to provide an intuitive understanding of the bias level, akin to recommendation labels on The Factual and NewsGuard. \textbf{Mechanism:} Reference-framed summary. \textbf{Cog Role:} Calibration and interpretable magnitude judgment.} \\\hline

 \parbox[t]{\linewidth}{ \textbf{Bias Highlights:} Visually shows which words are biased. (\Cref{fig:indicator_highlights}) \\  \textbf{Data Type:} Biased Words (in-text / local). \\  \textbf{Encoding:} \textit{Bias Highlights} marks biased words in place, using an in-text highlight that uses perceptual salience to direct attention to the specific linguistic cues that constitute bias. We create the highlights based on the expert annotations (s. supplementary material). The design is included due to its promising results in previous studies \cite{strobeltGuidelinesEffectiveUsage2016a, spindeHowWeRaise2022, simpleFramingInterventionBaumer2015}. \textbf{Mechanism:} Example-based learning. \textbf{Cog Role:} Direct localization of bias cues.} \\\hline

 \parbox[t]{\linewidth}{ \textbf{Political:} Displays the outlet's political slant. (\Cref{fig:indicator_pol})\\  \textbf{Data Type:} Slant of Outlet (1=left, 5 = right; source-level). \\ \textbf{Encoding:} Adapted from An et al. \cite{An_Cha_Gummadi_Crowcroft_Quercia_2021} and Spinde et al. \cite{spindeHowWeRaise2022}, \textit{Political} employs a five-level political scale (left, lean left, center, lean right, right). The classification is based on the AllSides Media Bias Chart.\footnote{\url{allsides.com/media-bias/media-bias-chart}} Despite conflicting research \cite{An_Cha_Gummadi_Crowcroft_Quercia_2021, spindeHowWeRaise2022}, political scales remain common on platforms like AllSides or The Factual. Inclusion choice prioritized ecological relevance over theoretical completeness. \textbf{Mechanism:} Context cue. \textbf{Cog Role:} External ideological framing.} \\\hline

\parbox[t]{\linewidth}{\textbf{Emotionality:} Displays what sentiments are in the statement. (\Cref{fig:indicator_emo})\\  \textbf{Data Type:} \% of positive/negative/neutral sentiments (text-level).\\  \textbf{Encoding:} The indicator explores the influence of emotional language on the perception and detection of linguistic bias. While emotionality is not synonymous with bias, there may be a strong correlation between the two, making it a crucial aspect to consider. The \textit{Emotionality} indicator visualizes the percentage of positive (white), negative (black), and neutral (grey) sentiments in the statement as a pragmatic operationalization of affective tone \cite{harrisSearchingDiversePerspectives}. To derive sentiments, we used Google's Natural Language API \cite{googlecloudnlp}. \textbf{Mechanism:} Affect cue. \textbf{Cog Role:} Emotional salience.} \\\hline

 \parbox[t]{\linewidth}{ \textbf{Trust:} Displays credibility of statement with single score/icon. (\Cref{fig:indicator_trust})\\ \textbf{Data Type:} Aggregated score from extremeness of political slant and percentage of biased words (text- + source-level). \\ \textbf{Encoding:} The \textit{Trust} indicator shows a visual scale with six shields, resembling a trust-based star rating system and consistent with research on icon-based trust cues  \cite{kenningSupportingCredibilityAssessment2018, kimSaysWhoEffects2019, NewsGuard2022}. It combines elements of both recommendation and information. As a mental shortcut, \textit{Trust} aims to provide a quick, intuitive understanding of the credibility of a news source or statement. The number of filled-out shields indicates the level of trustworthiness (zero = low, six = high). The score combines extremeness of the political rating of the source and the percentage of biased words in the statement 50/50. \textbf{Mechanism:} Heuristic credibility cue. \textbf{Cog Role:} Fast credibility judgment.} \\

\bottomrule
\end{tabular}
\end{table}

Three indicators focused directly on linguistic bias. \textit{Bias Bar} and \textit{Bias Gauge} encode the proportion of biased words in a statement as a summary cue, while \textit{Bias Highlights} marked biased words in the text as a local in-text cue. Grey highlighting was chosen over color to minimize evaluative connotations \cite{Ware2004InformationVisualizationPerception}.
Unlike the remaining indicators, which summarize information externally through scales, scores, or labels, \textit{Bias Highlights} localize evidence at the word level and may support learning through repeated exposure to examples of biased language \cite{strobeltGuidelinesEffectiveUsage2016a, Ware2004InformationVisualizationPerception}.
Three additional indicators presented summary cues commonly used in practice: \textit{Political} showed the political slant of the source on a five-level scale, \textit{Emotionality} showed the proportion of positive, negative, and neutral sentiment in the statement as a pragmatic approximation for emotional intensity through sentiment polarity by NLP tools, and \textit{Trust} summarized content credibility as a shield-based score.
The trust score is calculated and shown as shields from the average score of political rating on AllSides (-2 for far right/far left, -1 for right/left, 0 for center) and amount of biased words (0: <10\%, -1: <30\%,  -2: >30\%) as a symmetric baseline in the absence of validated weights.\footnote{Sensitivity checks with 25/75 and 75/25 splits correlated with the 50/50 baseline at r = .90 and r = .85 (Spearman), indicating the trust score was largely stable across weightings, though the 75/25 split produced somewhat greater divergence, suggesting that statements from outlets with high political extremeness are most sensitive to this choice.}
%> cat("r (25/75 vs 50/50):", r_25_vs_50, "\n")
%r (25/75 vs 50/50): 0.904 
%> cat("r (75/25 vs 50/50):", r_75_vs_50, "\n")
%r (75/25 vs 50/50): 0.85 

We selected these six indicators to achieve coverage across the most common data types used in deployed bias and credibility tools (see \Cref{sec:rw:design} and supplementary review table), rather than to construct a fully factorial design.
The comparison allowed us to test whether signals and interventions commonly used for credibility or veracity discernment also affect linguistic bias perception and detection.
We further ensured that the indicator data types could be automatically computed with classifiers for later implementation.
Each indicator represents a distinct data type: bias amount (\textit{Bias Bar}, \textit{Bias Gauge}), biased word locations (\textit{Bias Highlights}), political slant (\textit{Political}), sentiment composition (\textit{Emotionality}), and credibility (\textit{Trust}).
Including two bias-amount indicators (Bias Bar and Bias Gauge) was deliberate as they share the same underlying data but differ through the presence of a categorical reference frame established in \Cref{sec:method:material}, allowing a direct test of whether reference framing affects detection beyond raw magnitude encoding.
To improve comparability across conditions, all indicators used a grayscale design \cite{LixunTrustworthy2019} and were placed above the statement in similar size, except for \textit{Bias Highlights}, which appeared within the text.
This is not a fully controlled factorial manipulation, covers a limited amount of indicators, and the conditions differ along multiple dimensions.
We therefore treat this as a coverage study of deployed indicator types for linguistic bias rather than an extensive decomposition of individual encoding variables, and interpret effects at the indicator level rather than attributing them to specific encoding features.

\subsection{Measures}
\label{sec:meth:measures}
Our main dependent variables were bias detection and bias perception.

For detection, participants completed a word-level annotation task in which their selections were compared with the expert standard. We computed both F1 and \(d'\) scores. F1 captures the balance between precision and recall and is widely used in computer science studies on bias detection \cite{F1score, spinde2023media, Hinterreiter2024News}. \(d'\) is commonly used in psychology research and captures sensitivity in distinguishing biased from unbiased words, helping to separate discrimination ability from response bias \cite{bataillerSignalDetectionApproach2022a}.

For perception, we measured perceived bias after each statement using a standard bias item \cite{spindeYouThinkIt2021}, perceived emotionality \cite{RelianceMartel2020}, trust in the statement \cite{appelmanMeasuringMessageCredibility2016, kimCombatingFakeNews2019a}, and sharing intention using scales commonly used in misinformation research \cite{epsteinSocialMediaContext2023, claytonRealSolutionsFake2020, RelianceMartel2020}.
All questions are listed in \Cref{tab:questions}.

In addition, we recorded gender, age, education, political leaning, news consumption, topic opinion, and topic relevance as covariates.
Especially controlling for political leaning is important as articles aligning with readers' political orientation are perceived as less biased \cite{spindeHowWeRaise2022}, known as the congruency effect.
Therefore, we calculate political congruence for each statement.
Political leaning is measured on a scale from -10 (very liberal) to 10 (very conservative).
Further, we assess the personal relevance of the topic and corresponding opinion, following Moravec et al. \cite{sys1sys2InterventionsMoravec2020}.

\subsection{Materials}
\label{sec:method:material}
We created the study materials in four steps. First, we selected two politically contested topics prominent during data collection: Elon Musk’s Twitter/X acquisition and abortion. We collected 54 news articles and assigned each source a political ranking using the AllSides five-category system.
Second, we extracted 40–70 word statements that remained understandable without article context. Statements were preliminarily labeled as low, medium, or high bias with a political leaning. We used short statements to match short-form news exposure on social platforms and feeds.
Third, three researchers with prior experience in media bias annotation independently labeled biased words in the statements ($n = 1,134$), following the annotation guidelines of Spinde et al. \cite{spindeNeuralMediaBias2021} and definitions of linguistic bias \cite[Table 3]{spinde2023media}.
A word was considered biased if all three annotators agreed or, when two marked it, after discussion and consensus, following a procedure similar to Fuhr et al. \cite{fuhrInformationNutritionalLabel2018a}.

The inter-annotator agreement was Krippendorff's $\alpha = 0.47$, reflecting task subjectivity. Still, this exceeds prior word-level agreement and the sentence-level agreement reported by Spinde et al. \cite{spindeNeuralMediaBias2021}. The two most experienced annotators achieved $\alpha$ of $0.61$, indicating that experience improves agreement.

In total, 119 words were labeled as biased by all three annotators, while 97 additional words were labeled by two annotators and later resolved through discussion. For example, in the statement: \textit{"But now members of the anti-abortion movement are grappling with whether their greatest legal triumph led to political defeat in this week’s midterms. With historical trends and a curdling economy suggesting that Democrats could face historic losses, Team Blue made a risky bet on abortion rights."}, all annotators identified "grappling" and "Team Blue made a risky bet" as biased. Annotator 1 also labeled "on abortion rights," whereas Annotators 2 and 3 labeled "greatest legal triumph" and "curdling." After discussion, they agreed on "greatest legal triumph" and "curdling" as biased.

Fourth, we calculated the share of biased words in each statement, excluding stop words, and used these shares to define low-, medium-, and high-bias categories for \textit{Bias Gauge}.
Statements with 0--10\% biased words were labeled low bias, 10--30\% medium bias, and above 30\% high bias.\footnote{Shifting the low/medium threshold by ±3\% changed the category assignment of 1 statement (7\% threshold); shifting the medium/high threshold by ±5\% changed 3 statements (35\% threshold). Across all shifts, 11 of 15 statements retained their original category.} The final stimulus set contained 15 statements: nine for the indication phase and six for the testing phase, balanced across bias level and political slants.

\subsection{Procedure}
\label{sec:meth:procedure}
The study was implemented as a custom web application and followed three phases. In the \textit{intro} phase, participants were informed of the study's content and data processing and gave informed consent. Participants then completed demographics, and received a short explanation of linguistic bias with examples. A multiple-choice attention check ensured basic understanding of the concept of linguistic bias.
%todo ethics?

In the \textit{indication} phase, participants were randomly assigned to one of the six indicator conditions or the control condition. They then read nine short news statements on the Twitter/X topic. After each statement, they answered six questions on perceived bias, sentiment, trust, and sharing intention. We added a directed-response item as a second attention check to identify automated or inattentive responses.

In the \textit{testing} phase, we removed all indicators and presented six abortion-related statements. Participants were tasked to identify biased words by clicking on them and then answered the same perception questions as before.
This study design estimates whether representation effects transfer to new content and persist after cue removal, rather than only measuring immediate in-situ assistance.
%We removed the indicators in this phase to test whether prior exposure supported later bias detection without direct visual assistance and to reduce immediate anchoring effects.

At the end of the study, we asked participants if their input is usable for research to ensure data quality, with assurance of full pay independent of their answer.
Before the main study, a pre-test with identical setup and pay was conducted with 14 participants to identify and resolve any potential issues.
As an anonymous online survey via Prolific, formal ethics board review was not required under our institution's guidelines. The consent form all participants filled out prior to participation is available in the supplementary materials (\Cref{sec:supplemental_materials}).

\subsection{Statistical Analysis}
\label{sec:meth:stats}
We analyzed the effects of indicators on detection and perception using three linear mixed models in R\footnote{RStudio 2023.06.0+421 with lmerTest, lme4, psych, irr. LMMs are fit using the lme4 package with p-values and df calculated via the Satterthwaite approximation. Model comparisons are conducted using likelihood ratio tests with a chi-square distribution.} for F1, \(d'\) (RQ1), and perceived bias (RQ2). The fixed effects included indicator group, perceived emotionality, political leaning, political congruence (for liberals, moderates, and conservatives), age, gender, education, topic opinion, and topic relevance. We also modeled the interaction between political leaning and congruence (RQ4). Ground-truth statement bias and statement sentiment were also modeled as fixed effects to account for material differences. Participant ID was included as a random intercept.

To examine relationships among perceived bias, sentiment, trust, and sharing intention (RQ3), we computed Spearman correlations. We evaluated the internal consistency of the trust construct with Cronbach's \(\alpha\) and Bartlett's test of sphericity.  We conducted all analyses with a 95\% confidence level. For an initial power analysis, we used G*Power, targeting an .8 power to detect a medium effect size of $f = .25$, with a standard $\alpha$ error probability of .05, resulting in 225 required participants. We preregistered the study on OSF.\footnote{\url{osf.io/ekd6s/?view_only=e5dd63a6342448a5befd3c53fa3496a0}}

\subsection{Participants}
\label{sec:method:part}
We recruited 241 participants from the United States via Prolific and randomly assigned them to the seven study groups. Participants were excluded if they failed the initial attention check twice (\(n = 8\)), reported that their data should not be used for research (\(n = 7\)), or did not mark any word as biased despite later indicating that statements were biased (\(n = 12\)). The final sample comprised 214 participants with an achieved statistical power of .78.
The sample was balanced by gender (51\% women, 47\% men, 2\% diverse), had a mean age of 37.05 years (\(SD = 13.64\)), and reported high English proficiency. Participants were somewhat more liberal than the general U.S. population (\(M = -3.39\), \(SD = 5.7\), scale \(-10\) to \(10\)) \cite{gallup2024party}. The educational backgrounds of participants were higher than the average U.S. population \cite{uscensus}: 54\% held at least a four-year degree.
Average news consumption was about once per day. Participants generally select news sources aligned with their political orientations.
Participants identifying as centrist ($M = 2.32$, $SD = 0.80$) or conservative ($M = 2.95$, $SD = 0.84$) were more likely to consume news with a more liberal slant relative to their own political views.
A linear regression model predicting news consumption frequency showed age as the only significant predictor ($R^2 = .12$, $F(1, 212) = 28.19$, $p < .001$) with an 0.032 points increase for each additional year.

\section{Results}
\label{sec:results}

\subsection{Validation}
First, we validated the materials and constructs. Across all three linear mixed models, residual diagnostics did not indicate substantial violations of normality or homoscedasticity. 
Participants' perceived bias aligned with the expert-rated bias labels and the calculated proportion of biased words ($\beta = 0.43$, 95\% CI $[0.37, 0.50]$, $p < .001$; \Cref{tab:lmmBiasPerception}), as illustrated in \Cref{fig:statmentbias}. This result suggests that the selected statements reflected different levels of linguistic bias and that participants were generally able to perceive these differences.
The trust construct showed high internal consistency. Cronbach's \(\alpha\) ranged from .95 to .98, and Bartlett's test of sphericity was significant (\(\chi^2(3) = 617.50\), \(p < .001\)), supporting the use of the three-item trust measure in the following analyses.

\subsection{Bias Detection (RQ1)}
\begin{figure}[!ht]
\centering
 \includegraphics[width=\columnwidth, alt={Notched boxplot of F1 scores across seven treatment groups. Bias Gauge has the highest central tendency, followed by Bias Highlights and Trust, while Bias Bar is lowest and the control group is lower-middle; all groups show broad overlap and variability.}]{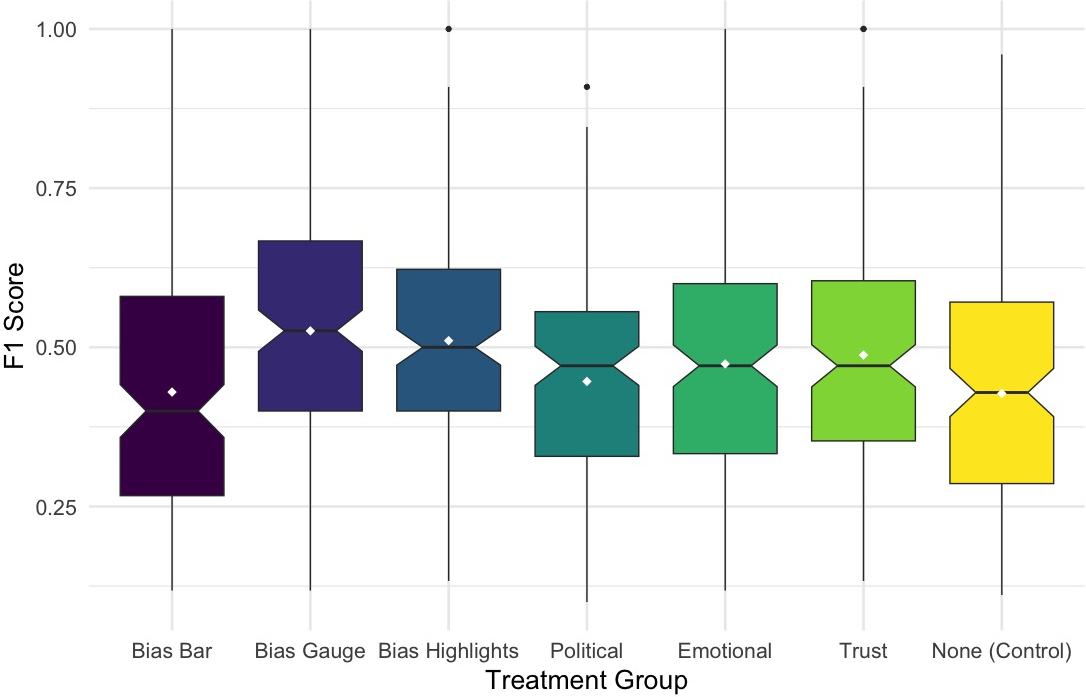}
 \caption{Boxplot with each group's F1 score measuring the degree to which participants detected bias compared to the expert standard. Higher scores indicate higher bias detection accuracy.}
 %Especially the lower F1 scores in the Control and Bar group and the higher F1 scores in Gauge and Highlights are visible.
 \label{fig:F1scatter}
\end{figure}

\begin{figure}[!htb]
\centering
\includegraphics[width=\columnwidth, alt={Notched boxplot of d-prime scores across seven treatment groups. Bias Highlights and Bias Gauge show the highest central values, Trust is somewhat elevated, Political and Emotional are intermediate, None is lower-middle, and Bias Bar is lowest; distributions overlap substantially and include several outliers.}]{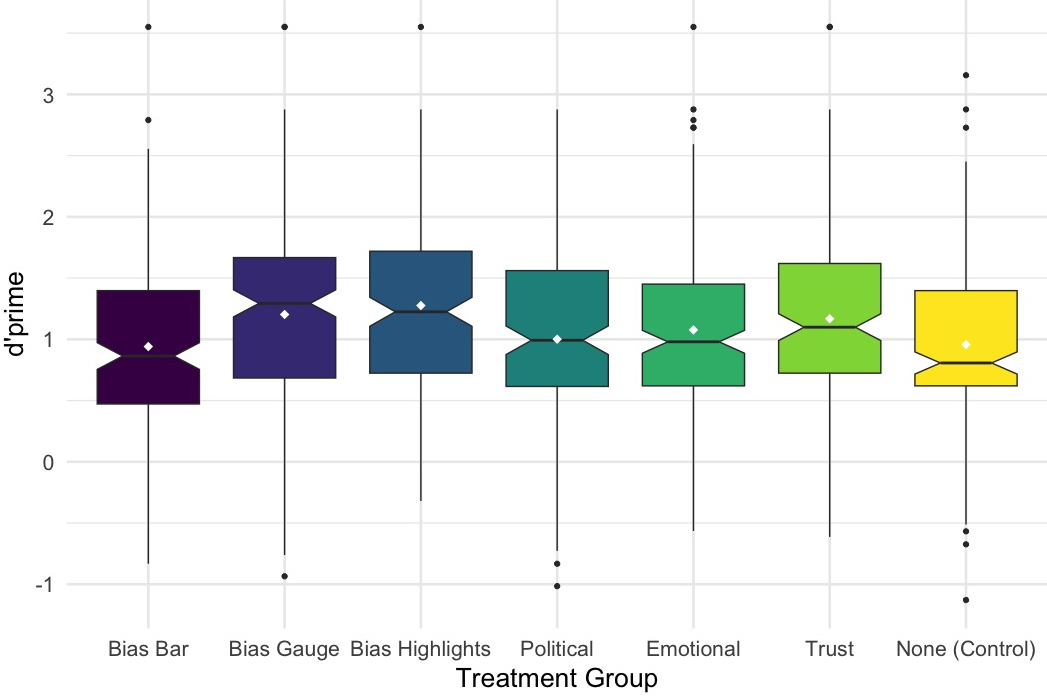}
 \caption{Boxplot with each group's $d'$ score measuring the degree to which participants discern bias compared to the expert standard. Higher scores indicate higher ability to discern bias.}
 %\Description{}
 \label{fig:dPrimescatter}
\end{figure}

To assess bias detection, we analyzed both F1 and \(d'\) scores from the word-level annotation task (\Cref{tab:meanF1dprime}). Mean scores were highest in the \textit{Bias Highlights} and \textit{Bias Gauge} conditions and lowest in the \textit{Control} and \textit{Bias Bar} conditions, visible in \Cref{fig:F1scatter} and \Cref{fig:dPrimescatter}.
\textit{Bias Highlights} and \textit{Bias Gauge} showed positive effects in the F1 model, with only \textit{Bias Highlights} showing positive effects in the \(d'\) model.

Across all participants, mean F1 was .474 (\(SD = 0.194\)), with recall of .32 and precision of .76. Participants therefore tended to avoid false positives more successfully than they detected all biased words.
The average $d'$ score was 1.08 ($SD = 0.452$), suggesting that while participants' ability to detect bias is limited, it is still above chance.
Participants with high $d'$ scores correctly identified 50-60\% of biased words with minimal false positives.
Overall, the subjective and fine-grained nature of bias makes it challenging for untrained participants to consistently identify bias \cite{spindeNeuralMediaBias2021, farberMultidimensionalDatasetBased2020b}, expectedly leading to lower scores.

\begin{table}[!htbp]
\footnotesize
\caption{Mean F1 and $d'$ scores of all indicator groups. Higher values indicate higher ability to detect and discern bias.}
{\begin{tabular}{l|c|c|c}
\cmidrule{1-4}
Indicator & F1 (SD) & $d'$ (SD) & Participants \\\hline
Bias Bar & 0.430 (0.203) & 0.941 (0.729) & 31\\
Gauge & 0.526 (0.193) & 1.203 (0.743) & 33\\
Highlights & 0.510 (0.186) & 1.276 (0.714) & 30\\
Political & 0.446 (0.181) & 1.001 (0.702) & 28\\
Emotionality & 0.474 (0.194) & 1.077 (0.740) & 33\\
Trust & 0.488 (0.184) & 1.168 (0.699) & 28\\
None (Control) & 0.427 (0.197) & 0.957 (0.679) & 31\\
\end{tabular}}
\label{tab:meanF1dprime}
\end{table}

\textbf{F1 Model.}
The F1 bias detection model (\Cref{tab:lmmf1}) explained 36\% of the variance overall (conditional \(R^2 = .36\), marginal \(R^2 = .13\)). Which indicator a participant saw explained a significant amount of additional variance in the F1 scores beyond predictors such as age, gender and education (\(\chi^2(6) = 12.75\), \(p = .047\)).
The strongest improvements were observed for \textit{Bias Highlights} ($\beta = 0.33$, $t(195) = 2.07$, $p = .040$) and \textit{Bias Gauge} ($\beta = 0.34$, $t(197) = 2.18$, $p = .031$), indicating that direct examples of biased language and contextualized bias amounts supported later detection better than the control condition.
The raw means of \textit{Bias Gauge} and control in \Cref{tab:meanF1dprime} corresponded to a model-adjusted gain of +0.067 F1 unit points (95\% CI [0.01, 0.13]), indicating a medium standardized effect ($\beta = 0.34$). \textit{Bias Highlights} produced a comparable adjusted gain of +0.067 F1 unit points (95\% CI [0.01, 0.12], $\beta = 0.33$). Both improvements were driven almost entirely by recall: \textit{Bias Gauge} raised recall from 0.25 to 0.38 while precision remained almost unchanged from 0.77 to 0.76. \textit{Bias Highlights} raised both recall from 0.25 to 0.37 and precision from 0.77 to 0.82, indicating the additional hits were accompanied by proportionally less false alarms. This observation points to a sensitivity gain, in line with a significant $d'$ effect below.
The confidence intervals for both indicators are relatively wide, suggesting the true difference to the control group may range from small to moderate.
Despite \textit{Bias Bar} using the same data type as \textit{Bias Gauge}, it showed no statistically reliable evidence of an effect. Neither did \textit{Emotionality}, \textit{Political}, nor \textit{Trust}.
Political factors showed small to large effects, detailed in \Cref{sec:res:pol}.

The F1 model showed a positive, small to medium effect between F1 scores and perceived bias. Sentiment in the statement had a small positive effect and bias in the statement had a small negative effect (\Cref{tab:lmmf1}).
%The model shows a positive, small to medium effect between F1 scores and perceived bias ($b = 0.037$, 95\% CI $[0.028, 0.045]$, $t(1029) = 8.24$, $p < .001$, $\beta = 0.26$) whereas sentiment in the statement has a small positive effect ($b = 0.052$, 95\% CI $[0.031, 0.074]$, $t(856) = 4.69$, $p < .001$, $\beta = 0.15$) and bias in the statement has a small negative effect ($b = -0.21$, 95\% CI $[-0.36, -0.065]$, $t(875) = -2.80$, $p = .005$, $\beta = -0.12$).
This means more sentiment in the statement and participants perceiving more bias lead to higher F1 scores. However, more bias in the statement decreased F1 scores.
Factors such as education and age did not impact detection accuracy.
Gender showed a non-significant, negative effect with men achieving slightly lower F1 scores than other genders (\Cref{tab:lmmf1}).

\textbf{\(d'\) Model.} The pattern for \(d'\) was similar.
The highest mean \(d'\) values occurred in the \textit{Bias Highlights} and \textit{Bias Gauge} groups. The model for $d'$ scores explained 42\% of the variance (conditional $R^2 = .42$), with 20\% attributable to the fixed effects (marginal $R^2 = .20$). Because $d'$ is expressed in standard-deviation units, the adjusted gains can be read directly as sensitivity improvements: \textit{Bias Highlights} increased $d'$ by 0.25 (95\% CI [0.05, 0.46], $\beta = 0.35$), about a quarter of a standard deviation over the control group. This suggests that the indicator group not only improved regarding annotation overlap with the expert standard but also sensitivity in distinguishing biased from unbiased words.
Additionally, the \textit{Trust} indicator showed a smaller, non-significant positive contrast of +0.19 $d'$ ($p = .055$) which we interpret as a tendency rather than an effect.
Neither the \textit{Bias Gauge}, \textit{Emotionality}, \textit{Bias Bar}, nor \textit{Political} indicators showed significant effects.

Similar to the F1 model, there was a positive relationship with a medium to large effect between perceived bias and $d'$ scores.
Again, higher bias in the statement had a strong negative relationship with $d'$ scores, with a small to medium effect (\Cref{tab:lmmd}).
A non-significant, negative relationship was observed between age and $d'$ score, aligning with existing literature \cite{littleFakeNewsSharingGuess2019}, with each additional year resulting in a slight decrease in $d'$ scores.
%  ($b = -0.004$, $p = .058$, $\beta = -0.08$)
Gender, specifically identifying as male, had a non-significant negative effect.
%($b = -0.1$, $p = .095$, $\beta = -0.14$).

Importantly, the Gauge pattern differed across metrics: it improved F1 but not $d'$. This suggests that the gauge may have calibrated the amount of bias participants marked, improving overlap with the expert annotations without clearly improving sensitivity in distinguishing biased from unbiased words. The interpretation matches the observed F1 increase, driven by higher recall with stable precision.
However, the limited or null effects for \textit{Trust}, \textit{Political}, and \textit{Emotionality} should be interpreted cautiously given the study’s slight underpowering.
Together, the two models indicate that indicators were most effective when they either showed biased words directly or placed bias amount into an interpretable context.

\subsection{Bias Perception (RQ2)}\label{sec:res:perception}
We next examined whether indicators influenced self-reported bias perception.
Participants' perceived bias increased with the ground-truth bias level of the statements, confirming that the selected material with its different bias levels was appropriate. However, the effect of indicators on perceived bias was weaker and less consistent than their effect on detection (\Cref{tab:lmmBiasPerception}).

The \textit{Bias Highlights} group showed a significantly higher perception of bias, a small effect significantly greater than the \textit{Control} group ($\beta = 0.19$, CI$[0.07, 0.74]$, $p = .024$).
Conversely, the \textit{Bias Bar} group recorded a significantly lower perception of bias across all statements, a small to medium effect ($\beta = -0.25 $, CI$[-0.88, -0.18]$, $p = .004$).
We discuss this further in \Cref{sec:dis:detection}.
Apart from \textit{Bias Highlights} and \textit{Bias Bar}, no other group showed significant effects.

\begin{figure}[!htb]
\centering
 \includegraphics[width=.97\columnwidth, alt={Scatterplot with overlaid trend lines showing the relationship between Perceived Bias on the horizontal axis and F1 on the vertical axis for all indicator groups and control. The x-axis is labeled “Perceived Bias” and spans roughly from about minus 3.5 on the left to plus 3.5 on the right. The y-axis is labeled “F1” and runs from near 0 at the bottom to 1.0 at the top, with labeled ticks around 0.25, 0.50, 0.75, and 1.00. Seven colored regression lines are drawn across the gray cloud, one for each indicator group listed in the legend on the right: Bias Bar, Bias Gauge, Bias Highlights, Political, Emotional, Trust, and None (Control). All lines slope upward from left to right, indicating a positive association between higher perceived bias and higher F1 for every group. Lines are similar, with Bias Gauge and Bias Highlights being the overall highest on the y axis and control having the steepest increase from low perceived bias to high perceived bias. The trust line is the most flat line. The background contains many small, semi-transparent gray points representing individual observations. Most points are concentrated on the right half of the figure, especially between perceived bias values of about 1 and 3, although some points also appear on the left side between about minus 3 and 0. The points are vertically dispersed at nearly every x-position, indicating substantial variability in F1 values. In the dense right-side cluster, many points fall between about 0.3 and 0.8 on F1, with some reaching close to 1.0 and others dropping nearer to 0.1 or 0.2.}]{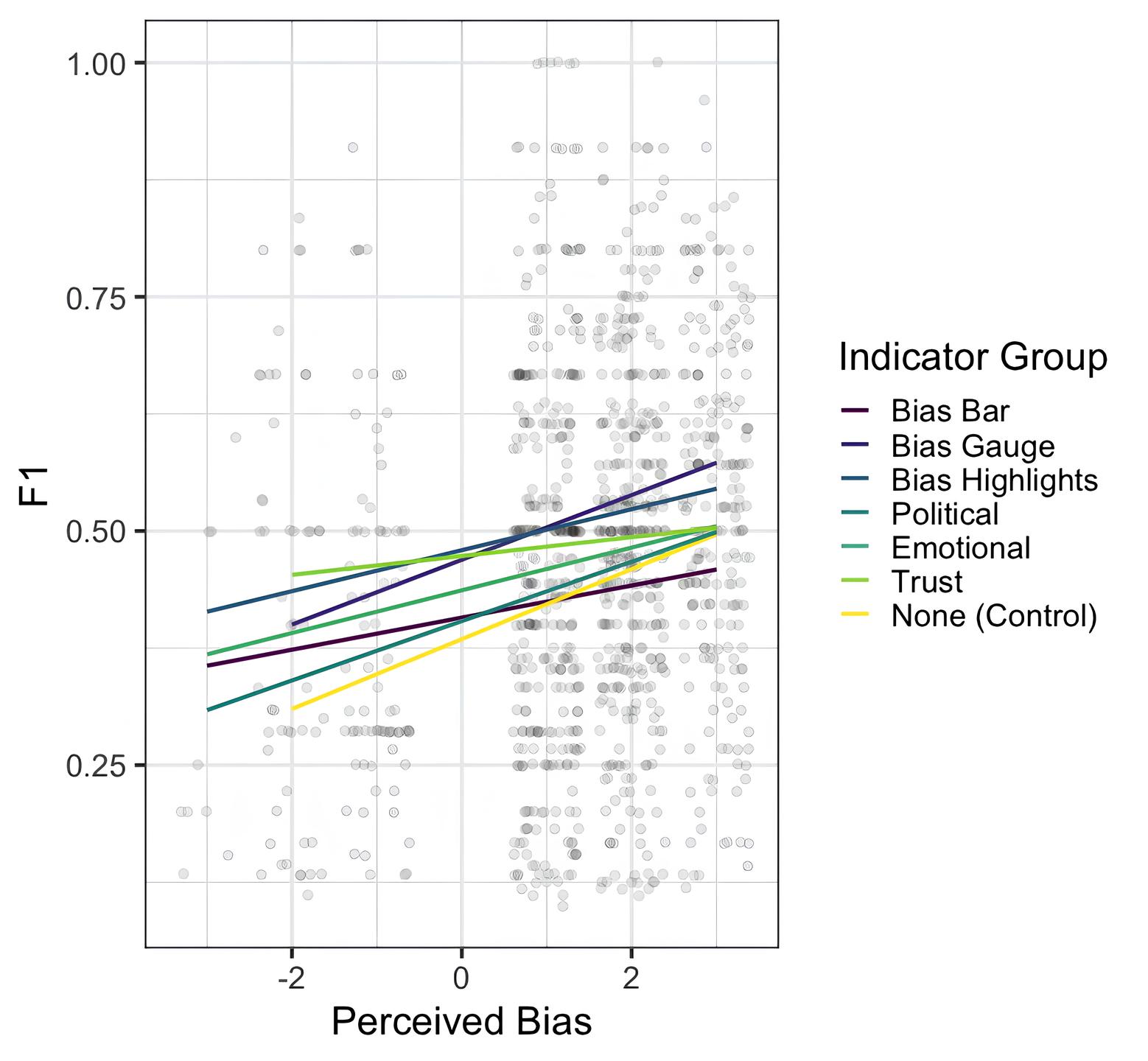}
 \caption{Perceived bias (question item rating bias in statement) vs. detected bias (F1 score) by indicator group. Higher F1 scores indicate higher bias detection accuracy.}
 %\Description{}
 \label{fig:f1vsperceived}
\end{figure}

\begin{figure}[!ht]
\centering
 \includegraphics[width=.97\columnwidth, alt={Scatterplot with overlaid trend lines showing the relationship between Perceived Bias on the horizontal axis and d′ Prime on the vertical axis for the same seven indicator groups. The x-axis is labeled “Perceived Bias” and again runs from roughly minus 3.5 to plus 3.5. The y-axis is labeled “d′ Prime” and spans from about minus 1 at the bottom to above 3 at the top, with major ticks at approximately minus 1, 0, 1, 2, and 3. As in the F1 plot, the background contains many semi-transparent gray dots representing individual observations. Most observations are clustered on the right side of the figure, especially between perceived bias values of about 1 and 3. These points are spread broadly in the vertical direction, with many d′ Prime values between about 0.5 and 2.2, some lower than 0, and some reaching above 3. The left side of the plot has fewer observations, mostly between perceived bias values of about minus 3 and minus 1, and those are also vertically dispersed. Seven colored trend lines are overlaid, corresponding to the legend on the right: Bias Bar, Bias Gauge, Bias Highlights, Political, Emotional, Trust, and None (Control). Every line slopes upward from left to right, showing that higher perceived bias is associated with higher d′ Prime for all indicator groups. Compared with the F1 figure, the upward trend is visually a bit more pronounced here because the lines span a larger vertical change across the plot. On the left side of the graph, when perceived bias is low, the lines lie close together around d′ Prime values of roughly 0.6 to 1.1. As perceived bias increases toward the right side, the lines rise to approximately 1.0 to 1.45. Bias Gauge and Bias Highlights appear to have the strongest positive slopes and end highest on the right side. Emotional and Political also rise clearly and end in the upper-middle range. Trust and None (Control) rise more modestly. Bias Bar appears among the flatter and lower lines, remaining near the bottom across much of the range.}]{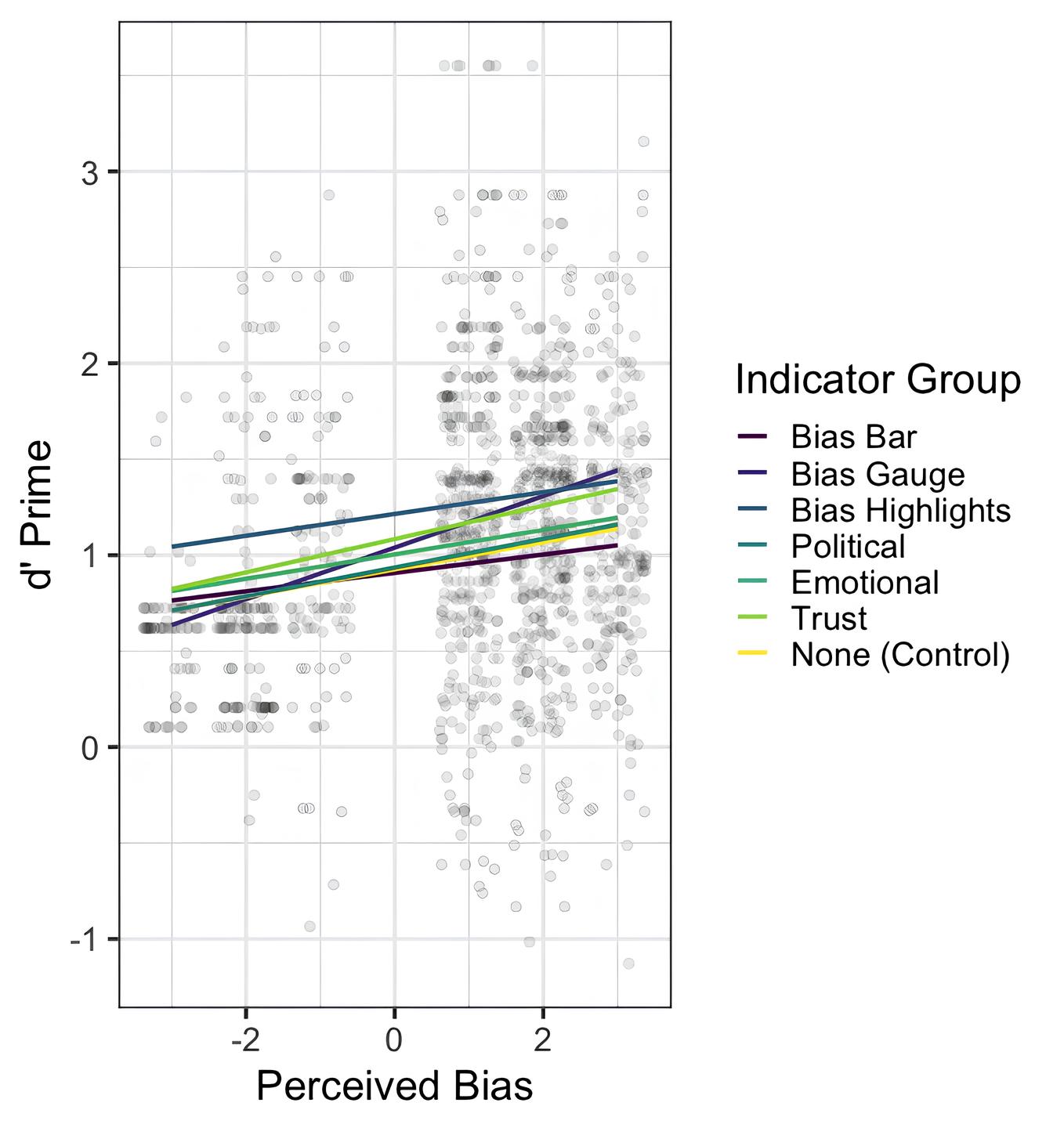}
 \caption{Perceived bias (question item rating bias in statement) vs. detected bias ($d'$ score) by indicator group. Higher $d'$ scores indicate higher bias discernment abilities.}
 %\Description{}
 \label{fig:dvsperceived}
\end{figure}

Overall, a higher perception of bias was associated with more accurate detection of bias, as evident in the \textit{Bias Gauge} group and visible in \Cref{fig:statmentbias} and \Cref{fig:f1vsperceived}.
In \Cref{fig:dvsperceived}, the \textit{Bias Gauge} group shows the highest decrease in $d'$ scores when perceived bias is low, yet records the highest $d'$ scores when perceived bias is high.
This suggests that when participants perceive less bias, their detection ability decreases.
Gender showed a small but significant effect, with men perceiving slightly less bias than other genders.

\subsection{Bias, Sentiment, Trust, and Sharing (RQ3)}
To examine how perceived bias relates to other judgments, we computed correlations between bias perception, perceived emotionality, trust, and sharing intention. We found that most factors were moderately to strongly correlated (\Cref{tab:corrMatrixIndicator}). The strongest relationship was between perceived emotionality and perceived bias ($\rho = .71$, $p < .001$), indicating that emotionally charged language was often interpreted as biased. This suggests that, in practice, readers might not separate linguistic bias from emotional tone, even though the two constructs are conceptually distinct.
Participants in the \textit{Emotionality} group rated statements as more emotionally charged ($M = 3.52$, $SD = 1.96$), whereas those in the \textit{Bias Bar} group perceived them as least emotional ($M = 2.88$, $SD = 1.73$).
Similarly, the F1 detection model showed a large effect of ground truth sentiment which the \(d'\) and perception models did not mirror. 

\begin{table} [!hbt]
\centering
\caption{Correlation of bias perception, sentiment perception, and willingness to share with statement bias, statement sentiment, and participants' trust (indication phase with indicator).}
{\begin{tabular}{l|c|c|c} 
\cmidrule{1-4}
 & Sentiment & Sharing & Trust\\\hline
Bias Perception   &.71*** & -.24*** &-.61***\\
Sentiment     & &-.14*** &-.48***\\
Sharing  & &  &.38***\\
\end{tabular}}
\vspace{2mm}
\par\footnotesize \textit{Note.} ‘***’: $\leq$ .001,  ‘**’: $\leq$ .01, ‘*’: $\leq$ .05, ‘.’: $\leq$ .1, ‘ ’: $>$ .1
\label{tab:corrMatrixIndicator}
\end{table}

Bias perception was also related to trust and sharing intention. Statements perceived as more biased tended to be judged as less trustworthy ($\rho = -.61$, $p < .001$) and less suitable for sharing ($\rho = -.24$, $p < .001$).
Trust was positively correlated with sharing ($\rho = .38$, $p < .001$).
%, such that a one-point increase in trust was associated with a 0.38-point increase in sharing likelihood ($\beta = 0.38$, 95\% CI [0.33, 0.42], $p < .001$).
%todo: could go in but I don't have the experimental models for trust and sharing and emotion in table form at hand rn
Political congruence was the strongest predictor of higher trust, and low-bias statements were trusted more than high-bias statements.
For example, liberal participants reported higher trust for liberal low-bias statements ($M = 5.04$) compared to liberal high-bias statements ($M = 4.12$).
Across experimental groups, the \textit{Trust} group reported the lowest average trust ($M = 3.80$, $SD = 1.48$), while the \textit{Political} group reported the highest trust ($M = 4.37$, $SD = 1.43$).

Sharing, transformed from the 7-point scale to -3 to 3, was lowest for highly biased statements (e.g., right-leaning, high-bias, $M = -2.51$, $SD = 1.04$) and highest for low-bias statements (e.g., centrist, low-bias, $M = -1.46$, $SD = 1.71$).
Participants in the \textit{Trust} group reported the lowest sharing intentions ($M = -2.21$, $SD = 1.35$), while the \textit{Bias Highlights} ($M = -1.63$, $SD = 1.56$) and the \textit{Control} ($M = -1.67$, $SD = 1.48$) groups report the highest.
The relationship between trust and sharing decisions connects to findings from misinformation research, where they are also correlated \cite{Chan02012026} and perceptions are shaped by quick, automatic judgments \cite{sys1sys2InterventionsMoravec2020}.

%Together, these results suggest that linguistic bias is embedded in a broader judgment process. Readers do not evaluate bias in isolation. Instead, they connect it to emotionality, credibility, and social sharing, which means that bias-mitigation tools may influence several downstream judgments at once.

\subsection{Political Congruency (RQ4)}
\label{sec:res:pol}
Overall, political leaning and congruency were strong factors influencing bias perception and detection, often stronger than the impact of an indicator.
%F1
In the F1 bias detection model (\Cref{tab:lmmf1}), participants' congruence with the statement and their political leanings' influence on F1 scores showed a medium effect ($\beta = 0.34$, CI$[0.02, 0.12]$, $p = .009$).
The \(d'\) detection model mirrored these findings (\Cref{tab:lmmd}). Here, congruency showed a significant small to medium effect ($\beta = 0.25$,  $t(1046) = 2.31$, CI$[0.03, 0.33]$, $p = .021$).
However, since congruency was coded across three levels (liberal, moderate, conservative), we focus our analysis on the interaction effects of liberal and conservative congruency, as these represent the most theoretically distinct positions.

While liberal participants achieved significantly higher F1 scores ($\beta = 0.34$, $p < .011$), the F1 model shows a large negative interaction effect between liberal participants and liberal statements ($\beta = -0.73$, $p < .001$), substantially larger than the non-significant conservative–congruency interaction ($\beta = -0.17$, $p = .411$).
%d'
The asymmetry was not visible in the \(d'\) model. Both liberals ($\beta = -0.39$, CI$[-0.53, -0.05]$, $p = .02$) and conservatives ($\beta = -0.38$, CI$[-0.52, -0.03]$, $p = .027$) showed significant medium negative interactions.
The \(d'\) model indicates that for conservative and more so for liberal participants, bias detection accuracy significantly decreased when statements align with their political beliefs.
%Perception Model
The perception model did not find general congruency effects (\Cref{tab:lmmBiasPerception}). However, liberal participants perceived significantly more bias in congruent statements compared to the baseline of moderates' congruency interaction ($\beta = 0.2$, CI$[0.01, 0.40]$, $p = .037$), diverging from the detection models.

Concluding, while some visual indicators affected perception, they did not override partisan bias and congruency effects.
Even effective indicators operate within a broader interpretive context shaped by prior beliefs, highlighting that user background is one of the main factors impacting bias perception and effectiveness of interventions.

\section{Discussion} \label{sec:dis}
This study examined how different visual indicators support the detection and perception of linguistic bias in short news statements and which other factors, such as political congruency, sentiment, and credibility, influence bias perception. The results show that indicators that support pattern learning through examples and direct cue localization \cite{healeyAttentionVisualMemory2012} (\textit{Bias Highlights}) or interpretable reference frames \cite{swellerCognitiveLoadDuring1988, franconeriScienceVisualData2021} (\textit{Bias Gauge}), improve linguistic bias detection more than abstract indicators. However, outcomes are more strongly influenced by political congruence.

The study further indicated a close connection between detected bias, perceived bias, actual bias in the statement, trust, sharing discernment, and perceived emotionality.
Their connection suggests that linguistic bias is embedded in a broader judgment process. Readers do not evaluate bias in isolation. Instead, they might connect it to emotionality, political stances, and credibility. Thus, representation choices may shape multiple downstream judgments beyond the main bias task.

We have also seen differences between bias perception and bias detection. For example, the \textit{Bias Bar} indicator did not impact detection but impacted perception. The difference matters for future tool designs. While detection and perception are related, they are not the same outcome. An indicator may help readers identify biased words without fully changing how biased they say a statement feels. For this reason, evaluating bias-mitigation tools requires both subjective perception measures and more objective test tasks.

Often, visual systems are evaluated as if information display alone can correct biased interpretation.
While our findings show that indicators can help readers notice one important layer of media bias, they do not address the broader problem of bias on their own. Linguistic bias indicators can be helpful but not sufficient. They can support awareness, yet operate inside social and cognitive contexts that influence what users are willing to notice or accept. Interpretation and impact on decision making remain strongly influenced by one's demographics, political worldview, the framing of the content, and prior information exposure.
Our work positions bias indicators not as standalone correctness tools, but as visual supports within a larger sensemaking process. 
Political worldview, emotional language, and trust judgments continue to shape how news is interpreted, even when visual support is provided.
From a visualization perspective, the question is therefore not only whether an indicator changes a judgment, but how it supports users in forming more reflective and better grounded judgments under real-world cognitive and political conditions.

%This study is among the first to assess word-level bias detection on short statements using both F1 and $d'$ scores, offering a more objective assessment, revealing discrepancies between what people think they see and what they can detect.

%These two designs support different cognitive processes. Summary indicators help answer: How biased is this text? Highlights help answer: Where is the bias? or What linguistic pattern constitutes bias?
%If participants repeatedly see examples of biased words during Phase 2, they may learn what linguistic bias looks like and transfer that knowledge to new statements.
%Text highlights exploit pre-attentive popout mechanisms that direct attention toward specific text regions and can improve later recall of highlighted information \cite{strobeltGuidelinesEffectiveUsage2016a, Ware2004Visu}. This suggests a potential mechanism through which repeated exposure to highlighted bias cues may support transfer to later unaided detection tasks.
%Text highlighting belongs to a broader class of embedded or in-text visual encodings that guide attention toward specific elements within a text stream. Unlike summary indicators that externalize information as aggregate scores or ratings, highlights localize evidence directly at the point of interpretation. Prior work suggests that such cues leverage pre-attentive popout and attentional guidance mechanisms, potentially supporting later recall and transfer \cite{strobeltGuidelinesEffectiveUsage2016a, Ware2004Visu}.

\subsection{Indicator Effects on Bias Metrics (RQ1, RQ2)} \label{sec:dis:detection}
In-text encodings that externalized the bias signal at the phrase level or translated quantitative values in an interpretable reference frame tended to raise bias detection more than abstract bias, trust, or sentiment summary encodings.
\textit{Bias Highlights} and, partly, \textit{Bias Gauge} produced the strongest improvements in word-level bias detection, whereas more abstract summary indicators with cues related to political leaning, sentiments and trust showed no or limited effects (RQ1, RQ2).
Information labels like \textit{Political} and \textit{Emotionality}, commonly used by news aggregators \cite{NewsGuard2022}, showed no significant effects on bias perception or detection.
While political classifications may be informative for curious readers, our study and prior work suggest that their value for increasing bias awareness is limited \cite{spindeHowWeRaise2022, parkNewsCubeDeliveringMultiple2009}. \textit{Trust} showed only minimal positive effects, and because the detection analyses were slightly underpowered, these findings should not be taken to mean that emotionality, political, or trust cues are ineffective in general.
However, abstract summary scores, a derived data type, require an additional mental step to map back to source content \cite{munznerVisualizationAnalysisDesign2014}. The mapping might be more fragile when prior beliefs are strong.
Overall, the findings are in line with previous research and suggest that indicators are most effective when they highlight concrete biased wording in-text \cite{strobeltGuidelinesEffectiveUsage2016a} or present overall bias with reference frames in an interpretable form \cite{swellerCognitiveLoadDuring1988, franconeriScienceVisualData2021}.

The two most effective indicators, \textit{Bias Highlights} and \textit{Bias Gauge}, rely on different mechanisms of support. \textit{Bias Highlights} likely help because they make the biased phrases explicit in-text. Despite gray highlights being a lower-salience choice \cite{strobeltGuidelinesEffectiveUsage2016a}, the effect suggests the mechanism is example-based learning  \cite{healeyAttentionVisualMemory2012} rather than perceptual salience alone.
\textit{Bias Gauge} may partially work because it provides a contextualized interpretation of bias amount which allows quick assessment through heuristics \cite{speierInfluenceInformationPresentation2006, sundarMAINModelHeuristic} linked to the example text.
In contrast, \textit{Bias Bar} may convey the same underlying information but be harder to interpret because it lacks a clear reference frame, consistent with previous work \cite{shahReviewGraphComprehension2002, swellerCognitiveLoadDuring1988, franconeriScienceVisualData2021}.
This may even reverse its effect on bias perception (\Cref{sec:res:perception}), as bars filled to only 30-50\% may anchor judgments around a summary score that readers interpret as 'not biased' when no explicit categories are shown.
The reference frame best suited for linguistic bias should be further investigated.

For \textit{Bias Gauge}, the significant F1 effect without a corresponding $d'$ effect or perception effect suggests that it influenced how much bias participants labeled rather than how accurately they identified it. The gauge may have served as a calibration cue that helped participants match the expected amount of bias rather than improving their ability to detect biased wording like \textit{Bias Highlights}.
The difference in cognitive support might support this, with \textit{Bias Highlights} providing explicit location of evidence, whereas \textit{Bias Gauge} operates at statement level, providing aggregate magnitude with context.

The differing effects observed in the bias perception model, compared to the detection models, suggest that noticing biased words and reporting a statement as biased are related but distinct processes. Detection can improve through exposure and learning, while explicit judgment remains more closely tied to prior attitudes and broader evaluative impressions. This reinforces the importance of evaluating visual interventions with both test and subjective perception measures rather than relying on a single measure alone.
% The two therefore measure different things — localized example-based learning vs. calibrated threshold matching — which limits direct comparison of effect sizes across them.

\subsection{Perceived Emotionality, Trust, and Sharing (RQ3)}
The strong correlation between perceived emotionality and perceived bias might suggest that participants do not strongly differentiate between emotionally charged language and linguistic bias. Dual-process theory could explain this conflation, as fast, affect-driven processing registers emotional tone as a signal of bias, while the slower reasoning needed to disentangle the two is rarely used during routine news reading \cite{moravecFakeNewsSocial2018}. Accordingly, participants in the \textit{Emotional} condition perceived statements as most emotionally charged, yet showed no gains in detection accuracy, suggesting that emotional salience amplified the conflation rather than resolving it. This implies that indicators showing emotional content may not reliably improve bias perception and detection.

The positive trust–sharing correlation replicates a documented pattern in misinformation research \cite{Majerczak2022}. Notably, participants in the \textit{Trust} condition reported the lowest trust and sharing intentions of any group.
The low trust suggests a backfire effect where explicit credibility cues might have triggered generalized skepticism rather than active assessment of bias \cite{Hoes2024}. Designers might want to treat such indicators with caution. Trust-based indicators may reduce engagement with legitimate content rather than improving critical evaluation of biased material \cite{NewsGuard2022}.

\subsection{Political and Demographic Factors (RQ4)}
Political leaning and political congruence were among the strongest factors across our models. Participants from the left or right tended to overlook bias in politically congruent statements but found them more trustworthy, regardless of their bias content. This tendency remained visible even when indicators were present, meaning that visual support did not neutralize partisan bias effects. Instead, indicator effects might have been filtered through readers' existing orientations.

%This finding is consistent with prior work on congruency effects \cite{spindeHowWeRaise2022, GarrettConservatives2021}, 
While Garrett and Bond \cite{GarrettConservatives2021} document greater conservative susceptibility to politically congruent but misleading information and Spinde et al. \cite{spindeHowWeRaise2022} reported stronger congruency effects regarding bias perception for conservatives, our results show a more nuanced pattern. Liberals perceived congruent statements as significantly more biased, yet their word-level detection accuracy dropped in one of the two models, suggesting they recognized bias as a general quality of the text but struggled to localize it at the word level.
We hypothesize that liberals may have a heightened general awareness of bias in congruent content, but this awareness does not translate into precise identification of biased language. Familiar, ideologically congruent phrasing may be harder for liberals to flag at the word level because it reads as natural, even when they consciously sense the text is biased overall. Conservatives, by contrast, showed no equivalent perception shift, suggesting that for them, congruency operates more automatically, reducing detection without triggering explicit bias awareness. However, as our sample contained more liberal participants, we cannot draw clear conclusions for conservatives.
Customized media literacy messages depending on the participant's political leaning and topic opinion from prior work \cite{vanDerMeerFighting2021} may be a strategy worth testing for media bias.
While demographic variables were less influential than political factors, their inclusion remains important as bias judgments are not made in a vacuum and should not be modeled as if users were interchangeable.

\subsection{Design Implications}

Prior work on bias and credibility indicators cluster around two broad principles: make evidence concrete and localized rather than abstract \cite{strobeltGuidelinesEffectiveUsage2016a}, and pair quantitative signals with interpretable reference frames \cite{swellerCognitiveLoadDuring1988, franconeriScienceVisualData2021}. Our results align with these principles but add empirical weight in the context of linguistic bias. \textit{Bias Highlights} confirms that in-text localization outperforms summary-level encodings for transfer-based detection, consistent with earlier work on highlights \cite{spindeHowWeRaise2022, simpleFramingInterventionBaumer2015, spindeEnablingNewsConsumers2020} as pre-attentive cues \cite{strobeltGuidelinesEffectiveUsage2016a}. \textit{Bias Gauge} confirms that categorical segmentation helps magnitude interpretation beyond raw bar encodings \cite{swellerCognitiveLoadDuring1988, franconeriScienceVisualData2021}, though its effect appeared in F1 but not d, suggesting calibration rather than genuine discrimination gains. However, effects of icon-based credibility cues \cite{sundarMAINModelHeuristic, kimSaysWhoEffects2019} do not clearly carry over to bias detection. \textit{Trust} showed only a marginal positive tendency, and its low sharing and trust scores across participants hint at backfire effects \cite{Hoes2024}.

We draw three design implications for bias mitigation tools from our results, summarized in \Cref{tab:predictorsEffects}.

\textbf{(1) Provide contextual learning opportunities.} Indicators that show examples or interpretable context outperform those showing only summary metrics. Designers should favor cues that help users understand \emph{why} content may be biased, not only \emph{how much}. Concretely, highlights, annotated examples, and interpretable scales are more promising than isolated scores or icons.

\textbf{(2) Design for political congruence effects.} Politically aligned content is consistently rated as less biased across the spectrum, so indicator designs must account for this asymmetry. Combining wording cues with comparative context (alternative viewpoints, source diversity, framing contrasts) and adapting indicators per target group \cite{vanDerMeerFighting2021} are two promising directions. Systems supporting easy cross-perspective comparison may further reduce congruence-driven blind spots.

\textbf{(3) Use bias cues, not sentiment cues.} Despite a strong correlation between perceived bias and perceived emotionality, the \textit{Emotionality} indicator did not directly impact bias perception. Tools should not treat emotional language as a proxy for bias and instead make the distinction between emotional tone and biased wording explicit and visible.

\begin{table}[!htb]
\caption{Predictors and their effects on detection (F1, $d'$) and perception (P.). ↑ positive, ↓ negative, ( ) non-significant effect.}
\footnotesize
{\begin{tabular}{>{\raggedright\arraybackslash}p{1.63cm}|>{\centering\arraybackslash}p{0.165cm}|>{\centering\arraybackslash}p{0.165cm}|>{\centering\arraybackslash}p{0.15cm}|>{\raggedright\arraybackslash}p{4.5cm}}
\cmidrule{1-5}
Predictor & F1 & $d'$ & P. & Result / Implication \\ \hline
\vspace{.5pt} Bias Bar & \vspace{.5pt} – & \vspace{.5pt} – &\vspace{.5pt} ↓ &\vspace{.5pt} Lowered bias perception significantly. Avoid designs without context frames. \vspace{1pt} \\
Bias Gauge & ↑ & – & – & Supported task, may provide reference frames for what is high/low bias. \vspace{1pt} \\
Bias Highlights & ↑ & ↑ & ↑ & Best supported task, examples might increase example-based learning. \vspace{1pt} \\
Political\linebreak Classification & – & – & – & While often used online, we found no effects.\linebreak \vspace{1pt} \\
Emotionality & – & – & – & While often used online, we found no effects. \vspace{1pt} \\
Trust Indicator & – & (↑) & – & Non-significant effect in the $d'$ model. Needs further research. \vspace{1pt} \\
Higher\linebreak Perceived Bias & ↑ & ↑ & – & Perceiving more bias was connected to detecting it more accurately. \linebreak \vspace{1pt} \\
Higher Bias in\linebreak Statement & ↓ & ↓ & ↑ & On statements with higher bias, detection accuracy decreased, but bias perception increased. \linebreak \vspace{1pt} \\
More Sentiment\linebreak in Statement & ↑ & – & ↑ & Only  in F1, not in $d'$. Needs further research. \linebreak \vspace{1pt} \\
Liberals & ↑ & – & – & Liberals detected bias more accurately, only significant in F1 LMM. \vspace{1pt} \\
Congruency\linebreak Liberals & ↓ & ↓ & ↑ & Decreased detection but increased perception of bias in congruent statements. \vspace{1pt} \\
Congruency\linebreak Conservatives & – & ↓ & – & Decreased detection in $d'$ model. Less represented in samples, more research needed. \vspace{1pt} \\
Gender (Male) & (↓) & (↓) & ↓ & Men detected and perceived less bias. \vspace{1pt} \\
Higher Age & – & (↓) & – & Higher age decreased bias detection in the $d'$ model (non-significant). \vspace{1pt} \\
\end{tabular}}
\label{tab:predictorsEffects}
\end{table}

%\textbf{(4) Support a multi-layered view of bias.} 

%These principles translate into several practical recommendations. First, bias-mitigation tools for everyday news consumption should prioritize interpretable cues over opaque summary scores. Second, systems should support comparison across content and viewpoints rather than presenting a single verdict. 

Linguistic cues address only one part of media bias. We suggest that future tools should connect word-level bias indicators with source-level, framing-level, and cross-article context. This layered approach would move beyond single labels toward richer visual analytic support for news interpretation.
Lastly, we argue that designers should treat bias indication as a reflective aid, not as an automated correction mechanism. The goal is not to replace the reader's own judgment, but to improve the conditions under which the judgment is formed.

\subsection{Limitations}
This study has several limitations.
We focused on short statements rather than full, multi-modal articles. This choice improved experimental control and reflected short-form news exposure, but it does not capture all forms of media bias, especially those that emerge through structure, omission, or article-level framing.
Material topic choices might also impact results. Because transfer was tested jointly with a material topic change, our results estimate transfer under changed material conditions rather than pure same-item retention.
Findings only generalize to first-exposure effects of indicator type on our  material format, but not to full articles, non-English contexts, or long-term use.

Researcher subjectivity, political stances, and Western-centric biases remain limitations, as no method can establish a fully objective gold standard for bias.
Although we achieved higher inter-annotator agreement than prior work on word-level bias annotation \cite{spindeNeuralMediaBias2021}, the task remains challenging and inherently subjective. Consequently, any ground truth is an approximation affecting evaluation.

The study tested immediate effects in a controlled setting and therefore cannot show how indicator use develops over longer periods of real-world news consumption.
The sample was somewhat more liberal and more educated than the general U.S. population.
Further, the study was slightly underpowered with small differences in group size.
We decided against resampling because news content can become outdated quickly, which could influence the results.

Including indicators based on bias, political leaning, sentiments, or trust reflects real-world practice, but it also complicates interpretation because they were originally developed for different goals.
Our indicators covered a range of commonly used constructs but differed in their visual encoding, data type, abstraction level, and the cognitive support they provide, e.g., \textit{Bias Highlights} exposed biased words in-text, while other indicators summarize bias at the statement level.
This limited the analysis of individual design features and may not be representative for the whole design space of bias indicators. Effects attributed to an indicator reflect the combined influence of its data type, visual encoding, cognitive support, and nature of the cue type or strategy rather than any single visual encoding.
While we standardized the designs for comparability, the differences may still have influenced participant responses. We therefore frame this study as exploratory rather than focused on individual encoding effects.

\section{Conclusion}

We compared six visual indicators for linguistic bias and found that indicators providing direct examples or interpretable context, especially \textit{Bias Highlights} and \textit{Bias Gauge}, best supported bias detection. 
Our results extend prior findings by showing that in‑text localization and categorical reference frames improve transfer‑based detection of linguistic bias while still interacting strongly with political congruence and perceived emotionality.
Political congruence remained a strong influence on both detection and perception, showing that visual indicators can support reflection but cannot solely remove broader interpretive biases.
Our study position linguistic bias indicators as one useful component of a larger bias mitigation strategy, highlighting the value of designing visual tools that support learning, comparison, and reflection rather than relying on summary scores or binary judgments.

%=======================================

%Please note that for the reporting of any experimental results that involve human participants you are \textbf{required} to ``include an explanation as to why such a review was not conducted. For research involving human subjects, authors shall also report that consent from the human subjects in the research was obtained or explain why consent was not obtained'' \cite[Section 8.1.1.E]{IEEEPublications2025}.
\section*{Supplemental Materials}
\label{sec:supplemental_materials}
All supplemental materials are available on \href{doi.org/10.5281/zenodo.19347505}{\texttt{Zenodo, DOI 10.5281/zenodo.19347505}}, released under a \href{https://creativecommons.org/licenses/by/4.0/}{CC BY 4.0 license}.
They include the R code, the study material, study screenshots, all indicator graphics, and the literature review table.

\section*{Figure Credits and Copyrights}
\label{sec:figure_credits}
\Cref{fig:examplesinterventions} and \Cref{fig:examplesotherindicators} contain screenshots of the original papers or linked websites. \Cref{fig:examplesinterventions}: Fuhr et al, 2018 \cite{fuhrInformationNutritionalLabel2018a}; Spinde et al, 2022 \cite{spindeHowWeRaise2022}; Diana and Stamper, 2022 \cite{DianaReducing2022}. \Cref{fig:examplesotherindicators}: Spinde et al, 2022 \cite{spindeHowWeRaise2022}; Tully et al, 2019 \cite{designingTully2019}, Garrett and Poulson, 2019 \cite{flaggingGarret2019}.
All figures remain under the authors' own copyright, with the permission to be used here. We also share them under a \href{https://creativecommons.org/licenses/by/4.0/}{CC BY 4.0 license}.

%% if specified like this the section will be omitted in review mode
\acknowledgments{
This work was supported by the \href{https://www.hss.de/}{Hanns-Seidel Foundation}, the German Federal Ministry of Education and Research (BMBF) through the \href{https://www.daad.de/en/}{DAAD} (German Academic Exchange Service), the Bavarian State Ministry for Digital Affairs in the project “XR Hub” (Grant A5-3822-2-16), and partially supported by JSPS KAKENHI Grant JP24H00732, by JST CREST Grants JPMJCR20D3 and JPMJCR2562 including AIP challenge program, and by JST K Program Grant JPMJKP24C2 Japan.
}

\bibliographystyle{abbrv-doi-hyperref}
\bibliography{template}

\appendix
\crefalias{section}{appendix} % this is to make sure that cleverref switches to referring to Appx. X from here on
\section{Appendix}

\begin{figure*}[]
  \centering
  \includegraphics[width=1.7\columnwidth, alt={The figure is arranged as a two-by-two grid of rectangular panels with pale yellow borders and bold titles at the top. The four panels are labeled “Warning Message,” “Presentation Manipulation,” “Educational Prompts,” and “Social Cues.” Each panel contains a different example of how systems might signal problems or encourage reflection when users encounter questionable content.
The top-left panel, “Warning Message,” is mostly text. It contains a cautionary notice telling readers to beware of biased news coverage, read consciously, and not be fooled. Beneath that is a longer explanatory paragraph defining media bias as non-neutral tonality and word choice in the news, and explaining that media bias can consciously or unconsciously create a narrow and one-sided point of view that affects public debate and decision making.
The top-right panel, “Presentation Manipulation,” shows a social-media post styled like a platform interface. A notice overlays the post stating that visibility is limited because the tweet may violate rules against hateful conduct, with a “View” option beside it. This example signals that the platform has intervened in how the content is displayed rather than directly classifying it as false or biased.
The bottom-left panel, “Educational Prompts,” shows a post from a media literacy organization above a large graphic reading “TRAIN YOUR BRAIN TO SPOT FAKE NEWS.” Beneath the slogan are three simple prompts: double-check the source, be aware of your reaction, and watch for red flags. This panel emphasizes general user education rather than item-specific warning or annotation.
The bottom-right panel, “Social Cues,” shows a post preview with a warning banner underneath it stating that the content is disputed by users like you. This cue does not explain the content directly, but instead indicates disagreement or skepticism from other people in the user’s social environment.}]{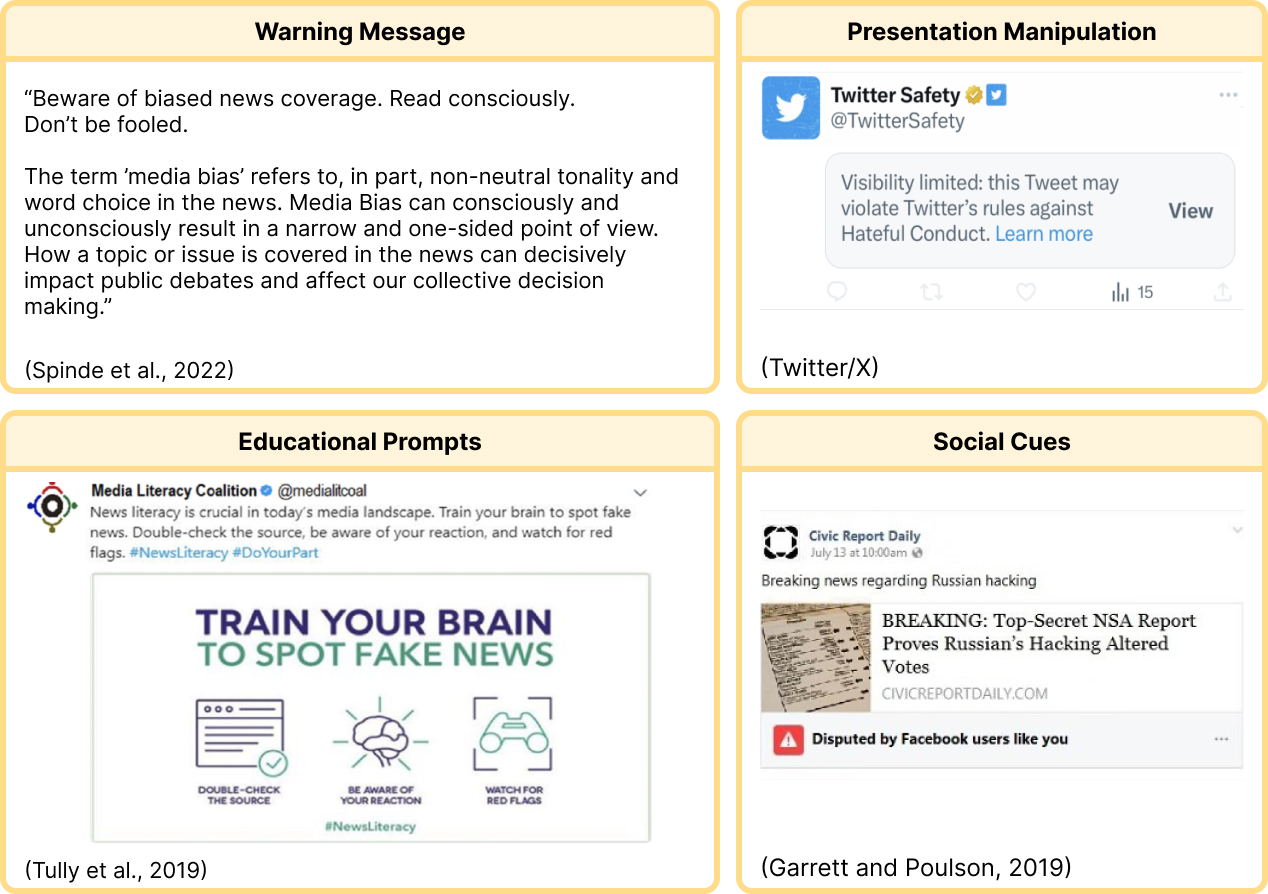}
  \caption{%
  	Examples for other indicator categories. \textbf{Warning Message}: Text warning about media bias before exposure to a biased article \cite{spindeHowWeRaise2022}. \textbf{Presentation Manipulation}: Limiting visibility of misinformation. \textbf{Education Prompts}: Prompting and explaining readers how to spot fake news \cite{designingTully2019}. \textbf{Social Cues}: Message warning that content was disputed by in-group peers \cite{flaggingGarret2019}.
  }
 \label{fig:examplesotherindicators}
\end{figure*}

\begin{figure*}[]
  \centering 
  \includegraphics[width=1.7\columnwidth, alt={The figure is a large collage divided into two main labeled regions. A purple-tinted header on the left reads “Information,” and a blue-tinted header on the right reads “Recommendation.” Under these headers are five boxed examples, each numbered and titled.
The first example, “(1) Information Nutrition Label,” resembles a nutrition facts label from food packaging. It uses a table format with rows and columns to summarize qualities of a news item, including categories such as fact, opinion, controversy, emotion, topicality, reading level, technicality, authority, and virality. Some rows show percentages or numeric values, and the design suggests a standardized breakdown of article characteristics.
The second example, “(2) Political Classification,” looks like a news card labeled “From the Left.” It includes a headline and an image thumbnail, followed by a small left-to-right political scale with multiple positions. The purpose appears to be classifying the political orientation of the article or source.
The third example, “(3) Bias Highlights with Explanation,” shows a text passage in which a phrase near the bottom is highlighted in yellow. A blue pop-up box overlays the text and defines “Subjective term” as language skewed by feeling, opinion, or taste. This example combines direct highlighting of biased wording with an explanatory tooltip.
On the right side, under “Recommendation,” the fourth example, “(4) Adapting Credibility Shield,” displays a green shield with a white checkmark next to a statement that the website generally maintains basic standards of accuracy and accountability. This design resembles a trust badge or approval marker intended to reassure readers.
The fifth example, “(5) Confirmation Bias Adaptive Prompt,” shows a headline with a colored scale beneath it and explanatory text warning that the headline may trigger bias in plausibility judgment. It describes a “Bias Danger: High” state and suggests the reader may overestimate plausibility because the content aligns with prior loyalties or attitudes. The visual framing implies a recommendation or reflective prompt rather than a direct content annotation.}]{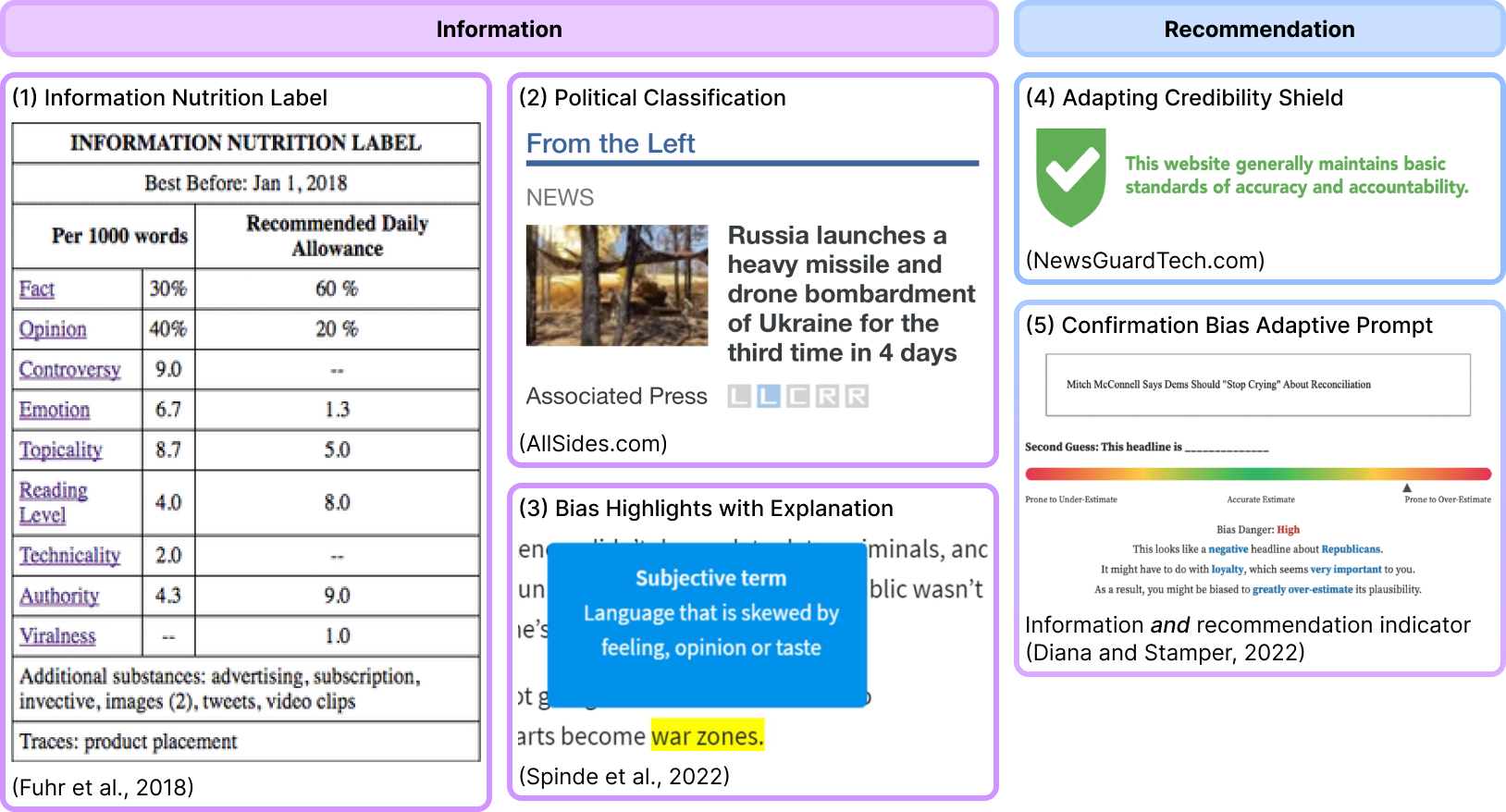}
  \caption{%
  	\textbf{Information Indicators}: (1) News Nutrition Label \cite{fuhrInformationNutritionalLabel2018a}: Displays key metrics about the article, such as its factual accuracy and level of opinion, helping readers assess its trustworthiness. (2) Political Classification: Shows the political leaning of the publishing outlet, giving context to the potential biases in the article. (3) Bias Highlights \cite{spindeHowWeRaise2022}: Marks biased phrases in the article in yellow. Hovering over the highlighted text provides an explanation of why the phrase is considered biased. \textbf{Recommendation Indicators}: (4) Credibility Shield: Uses color and icons to indicate how credible a news website is, providing visual cues through color and icons along with reasoning behind the assessment. (5) Adaptive Prompt for Confirmation Bias \cite{DianaReducing2022}: Combines both information and recommendation features by visually and textually displaying bias information while integrating personalized classifications that serve as recommendations, encouraging users to reconsider or reflect on their biases.
  }
 \label{fig:examplesinterventions}
\end{figure*}

\begin{figure*}[!ht]
\centering
 \includegraphics[width=\columnwidth, alt={A faceted boxplot-style figure showing how perceived bias varies across statements with different measured bias scores. The y-axis is labeled “Perceived Bias,” and the x-axis is labeled “Bias Score (\% of biased words in statement).” The y-axis uses three named levels: “Low Bias (-2)” near the bottom, “Medium Bias (0)” in the center, and “High Bias (2)” near the top. The x-axis presents nine separate statement conditions, each in its own narrow vertical panel, labeled from left to right: clot, llot, rlot, cmet, lmet, rmet, chit, lhit, and rhit. Beneath these panel labels, the corresponding bias-score percentages are 0\%, 8\%, 9\%, 12\%, 17\%, 27\%, 41\%, 53\%, and 69\%. A legend on the right titled “Statement Bias” maps dark blue to Low Bias, teal-green to Medium Bias, and yellow to High Bias.  The colored box shows the middle spread of responses, the horizontal line inside each box marks the median, the whiskers extend above and below the box, and a few small dark dots mark outliers in some panels. Visually, the boxes shift upward from left to right as the bias score increases, indicating that statements containing a higher percentage of biased words are generally judged as more biased by participants. Overall, the figure communicates a clear positive relationship between ground-truth bias score and perceived bias. As the percentage of biased words increases from 0\% on the far left to 69\% on the far right, the median perceived bias rises from clearly low, to mixed or moderate, to clearly high. The main visual pattern is a steady upward progression of the distributions across the nine panels, with some overlap and variability in the middle conditions but strong separation between the lowest-bias and highest-bias statements. Across the nine panels, perceived bias generally rises as the statement’s bias score increases: the first two low-bias conditions, clot (0\%) and llot (8\%), cluster low on the scale with medians around -2; the next two, rlot (9\%) and cmet (12\%), sit slightly higher with medians near -1 and a broader spread; the middle conditions, lmet (17\%) and rmet (27\%), shift upward to medians around 1 and show substantial variability; chit (41\%) appears transitional, also centered around 1 but not as high as the strongest bias conditions; and the final two panels, lhit (53\%) and rhit (69\%), sit highest overall with medians near 2, indicating that statements with the largest share of biased words were generally perceived as strongly biased, despite a few low outlier ratings.}]{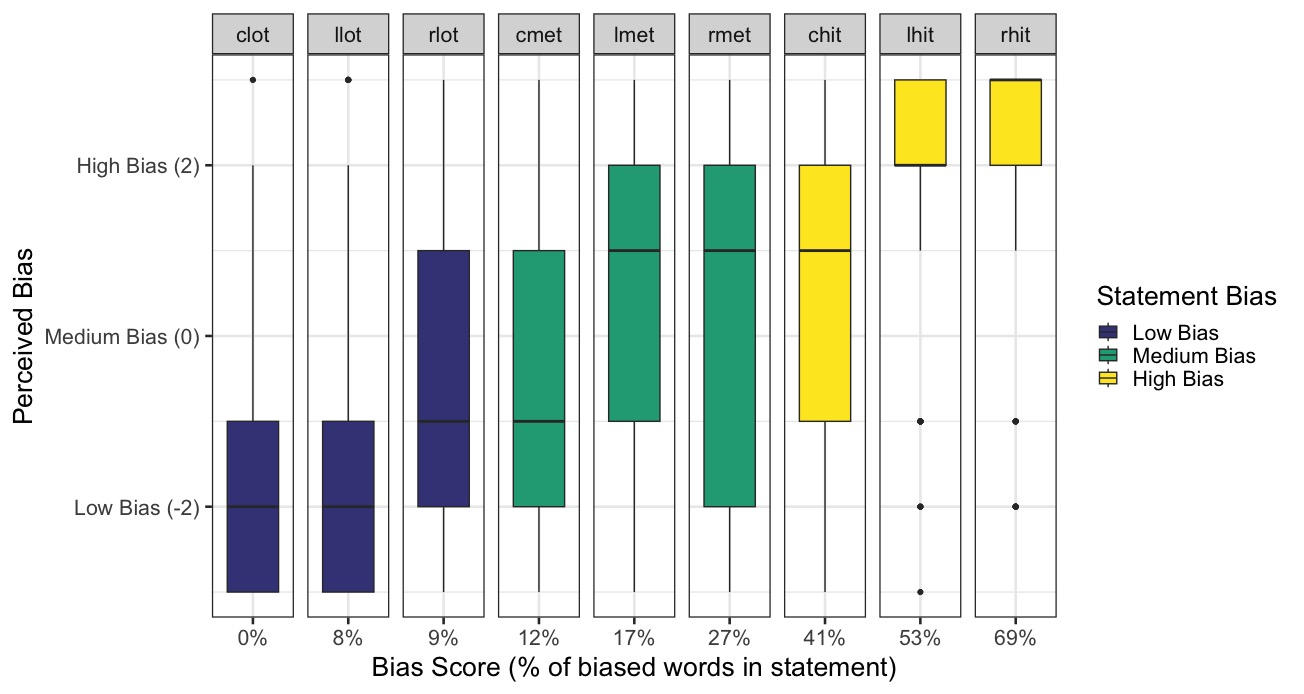}
 \caption{Perceived bias vs. percentage of bias for each statement in the indication phase. First letter: political slant of each statement (c = center, l = left/liberal, r = right/conservative), second and third letter = bias amount (lo = low, me = medium, hi = high), fourth letter = topic (t = Twitter/X).}
% \Description{}
 \label{fig:statmentbias}
\end{figure*}

\begin{table}[!htbp]
\footnotesize
\caption{Constructs and question items in the order that participants answer them after each statement.}
{\begin{tabular}{>{\raggedright\arraybackslash}p{0.125\linewidth}|>{\raggedright\arraybackslash}p{0.49\linewidth}|>{\raggedright\arraybackslash}p{0.24\linewidth}}
\cmidrule{1-3}
Construct & Question(s) & Scale  \\\hline
Bias & Please tell us what you think about the following sentence: In my opinion, this article is biased. \cite{spindeYouThinkIt2021, YeoEffect2019} & 'Strongly disagree' (1) to 'Strongly agree' (6) \\
Perceived Emo. & How emotional do you find the above content? \cite{RelianceMartel2020} & 'Not at all' (1) to 'Extremely' (7) \\
Trust & How (1. believable) (2. truthful) (3. credible) do you find the above content? \cite{appelmanMeasuringMessageCredibility2016, kimCombatingFakeNews2019a}& 'Not at all' (1) to 'Extremely' (7)
\\
Sharing & If you were to see the above article on social media, how likely would you be to share it? (on Facebook, WhatsApp, Telegram, or a similar platform). \cite{guessDigitalMediaLiteracy2020, epsteinSocialMediaContext2023} & 'Very unlikely' (1) to “Very likely” (7) \\
\end{tabular}}
\label{tab:questions}
\end{table}
%\section{Linear Mixed Models}
%F1 scores
\begin{table*}[h]
\centering
\footnotesize
\caption{Linear mixed-effects model for bias detection (F1 Scores) with participant as random intercept. Categorical predictors use the \textit{Control} group (Indicator) and Women (Gender) as reference levels. The indicators \textit{Bias Gauge} and \textit{Bias Highlights} are significant positive predictors.}
{\begin{tabular}{l|r|r|r|r|r|r|l}
\cmidrule{1-8}
 & \textit{b} & SE & $\beta$ & 95\% CI & \textit{t} & \textit{p} & Sig. \\\hline  
 Intercept & 0.37 & 0.05 & -0.24 & [0.28, 0.46] & 7.78 & $<$.001 & ***\\
Perceived Bias & 0.037 & 0.004 & 0.26 & [0.03, 0.05] & 8.24 & $<$.001 & ***\\
Liberal & 0.067 & 0.026 & 0.34 & [0.02, 0.12] & 2.55 & .011 & * \\
Conservative & -0.016 & 0.031 & -0.08 & [-0.08, 0.04] & -0.53 & .598 & \\
Congruency & 0.066 & 0.025 & 0.34 & [0.02, 0.12] & 2.60 & .009 & ** \\
\multicolumn{8}{l}{\textit{Indicator}} \\
\quad None (Control) & 0.00 & -- & -- & -- & -- & -- & (ref) \\
\quad Bias Bar & -0.012 & 0.032 & -0.06 & [-0.07, 0.05] & -0.36 & .721 & \\
\quad Bias Gauge & 0.067 & 0.031 & 0.34 & [0.01, 0.13] & 2.18 & .031 & * \\
\quad Bias Highlights & 0.064 & 0.031 & 0.33 & [0.01, 0.12] & 2.07 & .040 & * \\
\quad Political & 0.001 & 0.032 & 0.007 & [-0.06, 0.06] & 0.04 & .967 & \\
\quad Sentiment & 0.021 & 0.030 & 0.11 & [-0.04, 0.08] & 0.68 & .495 & \\
\quad Trust & 0.042 & 0.032 & 0.22 & [-0.02, 0.1] & 1.31 & .191 & \\
\multicolumn{8}{l}{\textit{Gender}} \\
\quad Women & 0.00 & -- & -- & -- & -- & -- & (ref) \\
\quad Men & -0.030 & 0.017 & -0.15 & [-0.06, 0] & -1.71 & .090 & . \\
\quad Other & 0.002 & 0.058 & 0.01 & [-0.11, 0.11] & 0.04 & .970 & \\
Age & -0.0004 & 0.0006 & -0.03 & [0, 0] & -0.71 & .476 & \\
Education & 0.003 & 0.006 & 0.02 & [-0.01, 0.01] & 0.53 & .598 & \\
Statement Bias & -0.21 & 0.076 & -0.12 & [-0.36, -0.07] & -2.80 & .005 & ** \\
Statement Sent. & 0.052 & 0.011 & 0.15 & [0.03, 0.07] & 4.69 & $<$.001 & *** \\
Topic Relevance & -0.003 & 0.003 & -0.05 & [-0.01, 0] & -1.18 & .239 & \\
Topic Opinion & 0.00003 & 0.004 & 0.0005 & [-0.01, 0.01] & 0.009 & .993 & \\
\multicolumn{8}{l}{\textit{Interactions}} \\
Lib. $\times$ Cong. & -0.14 & 0.038 & -0.73 & [-0.22, -0.07] & -3.71 & $<$.001 & *** \\
Cons. $\times$ Cong. & -0.033 & 0.040 & -0.17 & [-0.11, 0.05] & -0.82 & .411 & \\
\end{tabular}}
\vspace{2mm}
\par\footnotesize \textit{Note.} \textit{b} = unstandardized coefficient, $\beta$ = standardized coefficient. '***': \textit{p} $\leq$ .001,  '**': \textit{p} $\leq$ .01, '*': \textit{p} $\leq$ .05, '.': \textit{p} $\leq$ .1.
\label{tab:lmmf1}
\end{table*}

%d-prime scores
\begin{table*}[h]
\centering
\footnotesize
\caption{Linear mixed-effects model for bias detection ($d'$  Scores) with participant as random intercept. Categorical predictors use the \textit{Control} group (Indicator) and Women (Gender) as reference levels. The indicator \textit{Bias Highlights} is a significant positive predictor.}
{\begin{tabular}{l|r|r|r|r|r|r|l}
\cmidrule{1-8}
 & \textit{b} & SE & $\beta$ & 95\% CI & \textit{t} & \textit{p} & Sig. \\\hline  
 Intercept & 1.42 & 0.16 & -0.14 & [1.11, 1.73] & 8.82 & $<$.001 & ***\\
Perceived Bias & 0.14 & 0.01 & 0.34 & [0.12, 0.16] & 12.61 & $<$.001 & ***\\
Liberal & 0.14 & 0.09 & 0.20 & [-0.03, 0.32] & 1.61 & .110 & \\
Conservative & -0.15 & 0.11 & -0.20 & [-0.35, 0.06] & -1.35 & .179 & \\
Congruency & 0.18 & 0.08 & 0.25 & [0.03, 0.33] & 2.31 & .021 & * \\
\multicolumn{8}{l}{\textit{Indicator}} \\
\quad (None) Control & 0.00 & -- & -- & -- & -- & -- & (ref) \\
\quad Bias Bar & 0.04 & 0.11 & 0.05 & [-0.17, 0.25] & 0.34 & .738 & \\
\quad Bias Gauge & 0.14 & 0.11 & 0.19 & [-0.07, 0.34] & 1.28 & .202 & \\
\quad Bias Highlights & 0.25 & 0.11 & 0.35 & [0.05, 0.46] & 2.37 & .019 & * \\
\quad Political & -0.01 & 0.11 & -0.02 & [-0.22, 0.20] & -0.11 & .912 & \\
\quad Sentiment & 0.06 & 0.11 & 0.08 & [-0.14, 0.25] & 0.52 & .601 & \\
\quad Trust & 0.19 & 0.11 & 0.26 & [-0.02, 0.40] & 1.71 & .090 & . \\
\multicolumn{8}{l}{\textit{Gender}} \\
\quad Women & 0.00 & -- & -- & -- & -- & -- & (ref) \\
\quad Men & -0.10 & 0.06 & -0.14 & [-0.21, 0.01] & -1.68 & .095 & . \\
\quad Other & -0.04 & 0.20 & -0.06 & [-0.42, 0.34] & -0.21 & .832 & \\
Age & -0.004 & 0.002 & -0.08 & [-0.01, -0.00] & -1.91 & .058 & . \\
Education Score & 0.03 & 0.02 & 0.06 & [-0.01, 0.07] & 1.51 & .134 & \\
Statement Bias & -1.88 & 0.24 & -0.28 & [-2.36, -1.40] & -7.69 & $<$.001 & *** \\
Statement Sentiment & 0.05 & 0.04 & 0.04 & [-0.02, 0.12] & 1.40 & .162 & \\
Topic Relevance & -0.009 & 0.009 & -0.04 & [-0.03, 0.01] & -0.99 & .321 & \\
Topic Opinion & -0.013 & 0.013 & -0.06 & [-0.04, 0.01] & -1.03 & .305 & \\
\multicolumn{8}{l}{\textit{Interactions}} \\
Lib. $\times$ Cong. & -0.29 & 0.12 & -0.39 & [-0.53, -0.05] & -2.34 & .020 & * \\
Cons. $\times$ Cong. & -0.28 & 0.13 & -0.38 & [-0.52, -0.03] & -2.22 & .027 & * \\
\end{tabular}}
\vspace{2mm}
\par\footnotesize \textit{Note.} \textit{b} = unstandardized coefficient, $\beta$ = standardized coefficient. '***': \textit{p} $\leq$ .001,  '**': \textit{p} $\leq$ .01, '*': \textit{p} $\leq$ .05, '.': \textit{p} $\leq$ .1.
\label{tab:lmmd}
\end{table*}

%Bias perception
\begin{table*}[!ht]
\centering
\footnotesize
\caption{Linear mixed-effects model for bias perception with participant as random intercept. Categorical predictors use the \textit{Control} group (Indicator) and Women (Gender) as reference levels. \textit{Bias Highlights} was a significant positive predictor, whereas \textit{Bias Bar} was a significant negative predictor.}
{\begin{tabular}{l|r|r|r|r|r|r|l}
\cmidrule{1-8}
 & \textit{b} & SE & $\beta$ & 95\% CI & \textit{t} & \textit{p} & Sig. \\\hline  
Intercept & -2.21 & 0.23 & 0.05 & [-2.64, -1.78] & -9.77 & $<$.001 & ***\\
Statement Bias & 0.35 & 0.03 & 0.43 & [0.37, 0.50] & 13.51 & $<$.001 & ***\\
Statement Sentiment & 0.67 & 0.11 & 0.2 & [0.14, 0.26] & 6.39 &$<$.001 & ***\\
\multicolumn{8}{l}{\textit{Indicator}} \\
\quad (None) Control & 0.00 & -- & -- & -- & -- & -- & (ref) \\
\quad Bias Bar & -0.53 & 0.18 & -0.25 & [-0.88, -0.18] & -2.89 & .004 & ** \\
\quad Bias Gauge & -0.003 & 0.18 & -0.001 & [-0.34, 0.33] & -0.02 & .987 & \\
\quad Bias Highlights & 0.40 & 0.18 & 0.19 & [0.07, 0.74] & 2.28 & .024 & * \\
\quad Political & -0.05 & 0.18 & -0.02 & [-0.39, 0.30] & -0.25 & .804 & \\
\quad Sentiment & 0.006 & 0.17 & 0.003 & [-0.33, 0.34] & 0.04 & .972 & \\
\quad Trust & -0.10 & 0.18 & -0.05 & [-0.45, 0.25] & -0.54 & .593 & \\
Education Score & -0.01 & 0.03 & -0.007 & [-0.07, 0.05] & -0.30 & .764 & \\
Liberal & -0.12 & 0.15 & -0.06 & [-0.40, 0.16] & -0.84 & .402 & \\
Conservative & 0.15 & 0.19 & 0.07 & [-0.10, 0.25] & 0.81 & .420 & \\
Congruency & -0.11 & 0.15 & -0.05 & [-0.20, 0.09] & -0.71 & .477 & \\
\multicolumn{8}{l}{\textit{Gender}} \\
\quad Women & 0.00 & -- & -- & -- & -- & -- & (ref) \\
\quad Men & -0.20 & 0.10 & -0.09 & [-0.39, -0.01] & -1.99 & .049 & * \\
\quad Other & -0.07 & 0.33 & -0.03 & [-0.70, 0.56] & -0.22 & .825 & \\
Age & 0.0006 & 0.004 & 0.004 & [-0.01, 0.01] & 0.16 & .870 & \\
Topic Relevance & -0.017 & 0.015 & -0.03 & [-0.05, 0.01] & -1.07 & .286 & \\
Topic Opinion & 0.016 & 0.021 & 0.02 & [-0.02, 0.06] & 0.77 & .441 & \\
\multicolumn{8}{l}{\textit{Interactions}} \\
Lib. $\times$ Cong. & 0.43 & 0.21 & 0.2 & [0.01, 0.40] & 2.09 & .037 & * \\
Cons. $\times$ Cong. & -0.03 & 0.27 & -0.01 & [-0.27, 0.24] & -0.10 & .918 & \\
\end{tabular}}
\vspace{2mm}
\par\footnotesize \textit{Note.} \textit{b} = unstandardized coefficient, $\beta$ = standardized coefficient. '***': \textit{p} $\leq$ .001,  '**': \textit{p} $\leq$ .01, '*': \textit{p} $\leq$ .05.
\label{tab:lmmBiasPerception}
\end{table*}

\end{document}